\documentstyle [aps,pre,epsf,multicol]{revtex}

\begin{document}

\title{Macroscopic car condensation in a parking garage}
\author{Meesoon Ha and Marcel den Nijs}
\address{Department of Physics, University of Washington, Seattle, Washington 98195, USA}
\date{\today}

\maketitle
\begin{abstract}

An asymmetric exclusion process type process,
where cars move forward along a closed road that starts and
terminates at a parking garage, displays dynamic phase transitions into
two types of condensate phases where the garage becomes macroscopically 
occupied. The total car density $\rho_o$ and the exit probability
$\alpha$ from the garage are the two control parameters.
At the transition, the number of parked cars $N_p$ diverges
in both cases, with the length of the road $N_s$,
as $N_p\sim N_s^{y_p}$ with $y_p=1/2$.
Towards the transition, 
the number of parked cars vanishes
as $N_p\sim \epsilon^\beta$ with $\beta=1$, 
$\epsilon=|\alpha -\alpha^*|$ or $\epsilon=|\rho^*_o -\rho_o|$,
being the distance from the transition.
The transition into the normal phase represents also the onset of 
transmission of information through the garage. This gives rise to
unusual parked car autocorrelations
and car density profiles near the garage, which depend 
strongly on the group velocity of the fluctuations along the road.

\end{abstract}
\draft

\begin{multicols}{2}

\narrowtext

\section{Introduction}
\label{intro}

The scaling properties of nonequilibrium-type phase transitions are a 
topic
of intensive current research~\cite{Liggett,B-Stanley,Privman,M-Dickman}.
From the theoretical side, focus has been mostly on model calculations
using stochastic dynamics with simple local rules and local interactions.
These models serve as prototypes for various physical processes, such as
surface catalysis~\cite{CATALYSIS}, population growth~\cite{POPULATION},
surface growth~\cite{B-Stanley,RBDetc},
electronic transport in wires~\cite{TRANSPORT},
traffic flow~\cite{TRAFFIC}, and avalanches in granular
materials\cite{SOC}.
The simplicity of the models is justifiable
when, as we  expect, the scale invariant  collective fluctuations
at large length scales are universal and insensitive to most 
microscopic details.

Queuing phenomena,
like traffic jams behind slow moving trucks on single-lane highways,
are common to many nonequilibrium processes.
One of the interesting phenomena in such systems is the
transition from a finite queue to an infinitely long one, i.e.,
the transition from a queue that does not scale
with the road length $N_s$ to one whose length is proportional to $N_s$.
Such transitions can be induced by changing 
the total density of cars in the system or
by varying the probability rate at which cars pass the
truck~\cite{J-Lebowitz}.

A simplification of the above is first to replace the slow moving 
truck by a stationary object, and then to ignore the excluded volume 
effects in the queue by allowing the cars behind this stationary impurity 
to pile on top of each other. This bare-bone version of the phenomenon 
can then be rephrased in terms of the parking garage model 
introduced in this paper.
An asymmetric exclusion process (AEP) on a road starting and terminating 
at a garage with an infinite parking capacity.
The control parameters are the total number of cars, $N_c$,
in the system (which compared to the road length 
$N_s$, defines a total car density $\rho_o=N_c/N_s$) and
the probability $\alpha$ with which cars exit the garage 
(compared to the hopping probability along the road, set equal to one).
The infinitely long queue is represented by a macroscopic occupation of
the garage. The density of parked cars, $\rho_p=N_p/N_s$, 
is nonzero when the number of cars at the garage $N_p$ 
is proportional to $N_s$.

The analogy with equilibrium Bose condensation is obvious
as well as the basic question, i.e., whether
the scaling properties of such queuing transitions are universal.
For example, how sensitive are the
exponents $\beta$ and $x_p$ in
\begin{equation}
\rho_p\sim |\alpha-\alpha_c|^\beta  ~~~~{\rm for }~~ \alpha < \alpha_c
\end{equation}
and
\begin{equation}
\rho_p \sim N_s^{-x_\rho}  ~~~~{\rm at }~~ \alpha_c
\end{equation}
to the details of the queuing dynamics, particularly
to the simplifications we made.
As a start, we need to establish and understand the scaling properties
of the most simplest model in as much detail as possible.
The scaling properties of the parked car condensation transitions
presented here are indeed already surprisingly rich.
After this, we will be well positioned to
restore the omitted queuing details one-by-one, and
establish how robust those scaling properties are.

This paper is organized as follows.
The parking garage model is introduced in Sec.~\ref{model},
and the phase diagram is presented in Sec.~\ref{phasediagram}.
In Sec.~\ref{KPZ}, we explore the relation to 
Kardar-Parisi-Zhang (KPZ) type surface growth.
The group velocity is introduced in Sec.~\ref{groupvelocity},
and its important role in setting the density profile 
in the various phases is reviewed in Sec.~\ref{densityprofile}.
In Sec.~\ref{correlations}, we do the same for the density-density 
correlations. Sections~\ref{fsscaling} and \ref{carfluctuations} form the 
core of this paper.
We present our numerical Monte Carlo analysis for the parked car density
at the condensate-normal phase transitions in Sec.~\ref{fsscaling}, 
together with explanations of these results. 
Next, we do the same for the parked car fluctuations 
in various phases and at the phase boundaries 
in Sec.~\ref{carfluctuations}.
Finally, in Sec.~\ref{summary}, we summarize our results.

\section{Parking Garage Model}
\label{model}

Consider a one-dimensional road of $N_s$ sites
with periodic boundary conditions (PBC).
Each site on this road, $1\le x\le N_s-1$, is either empty or occupied
by only one car, $n_x=0,1$. The parking garage at site $N_s$ has no 
occupation upper limit, $N_p\equiv n_{N_s}=0,1,2,3,...$.
The total number of cars ($N_c$) in the systems is conserved.
From the perspective of other physical processes,
the cars can be reinterpreted just as well as, e.g.,
molecules (driven down a circular tube),
kinks in steps on crystal surfaces,
electrons driven around a wire by an electric field,
or domain walls in magnetic spin chains.

The stochastic update rule, from time $t$ to $t+1$ is sequential.
One of the sites, $1\leq x \leq N_s$, is selected at random
with uniform probability $p(x)=1/N_s$.
If that site is part of the road, $1\le x \le N_s-2$, and occupied,
the car moves forward to over one unit, presuming site $x+1$ is empty;
otherwise nothing happens.
If the selected site is the one just in front of the garage, $N_s-1$,
and is occupied, then that car moves with certainty into the garage,
irrespective of the occupation level of the garage.
If the chosen site is the garage, $x=N_s$,
the probability for a car to jump out of it onto site $x=1$
is equal to $\alpha$, independent of the number of cars at the garage,
provided that there is at least one available and site $x=1$ is empty.
This exit probability from the garage
is smaller than the hopping probability on the road, 
$0\leq \alpha \leq 1$.

The above process has two control parameters:
the total density of cars $\rho_o$ and
the probability $\alpha$ with which cars can escape from the garage.
We define the following quantities:
the total car density $\rho_o=N_c/N_s$,
the parked car density $\rho_p=N_p/N_s$,
and the on-the-road density $\rho_r=\rho_o-\rho_p$.
Only $\rho_p$ and $\rho_r$ fluctuate.

The parking garage is the new aspect to this otherwise well-studied
AEP. The latter is exactly soluble by 
the Bethe ansatz for periodic boundary conditions
(a closed loop road without garage)~\cite{G-Spohn},
and the stationary state properties are exactly known 
for an open road with reservoirs on both sides~\cite{Derrida}.
We will use and comment on these different setups and exact results 
in the following sections.

\section{The Phase diagram}
\label{phasediagram}

\begin{figure}
\centerline{\epsfxsize=7cm \epsfbox{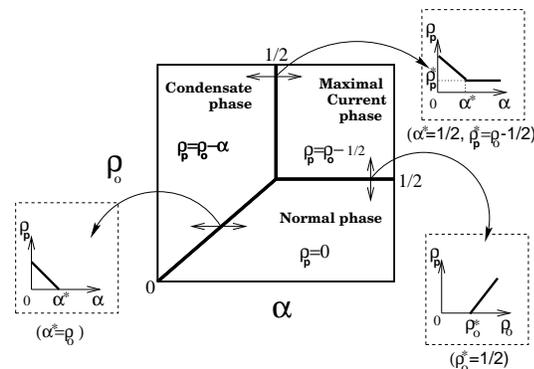}}
\caption
{Phase diagram with the total car density $\rho_o$ and
the exit probability $\alpha$ from the garage.
At the phase transition lines into the $N$ phase,
the parked car density $\rho_p$ vanishes, and
the garage seizes to be macroscopically occupied
(shown schematically in the insets).}
\label{newPD}
\end{figure}

Figure~\ref{newPD} shows our phase diagram.
It contains  three phases. In two of them the garage is 
macroscopically occupied: the condensate ($C$) phase 
(where the garage controls the density of cars on the road), 
and the maximal current (MC) phase 
(where the road capacity controls the flow).
In the normal ($N$) phase, the garage contains
a finite number of parked cars (typically only a few).
This structure can be deduced almost completely from earlier results 
for AEPs without the garage, such as the AEP 
with closed periodic boundary conditions (no garage), 
i.e., KPZ growth~\cite{G-Spohn,K-P-Zhang},
and the AEP with open boundary conditions 
hooked-up to reservoirs on both ends~\cite{Derrida,Krug}.

In the two condensate phases, $C$ and MC, 
the garage acts like a reservoir,
and the model reduces to the open road version 
studied by Derrida {\em et al.}~\cite{Derrida} and others~\cite{Krug}.
In their version, the road is in contact with car reservoirs 
at both ends, $x=1$ and $x=N_s-1$, 
such that a car jumps onto $x=1$ with probability $\alpha$ 
(if empty) and leaves from $x=N_s-1$ with probability $\beta$ 
(if occupied). In our model $\beta$ is always equal to one.

A dynamic second-order phase transition takes place between the MC phase 
(where the road density is at a constant value, $\rho_r=\frac{1}{2}$) and 
the $C$ phase (where the road density $\rho_r=\alpha$ varies with 
the inlet probability $\alpha$ and this result is already correct 
within mean field theory~\cite{Derrida,Krug}).
The MC phase appears because raising the density any further would 
reduce the flow efficiency (due to overcrowding).
This phenomenon has been canonized recently into 
a ``maximal current principle"~\cite{Popkov-S},
which states that the bulk road density takes the value 
that maximizes the flow.

Reduce the total number of cars inside the $C$ and MC phases,
i.e., walk in phase diagram towards the $N$ phase 
along a line of constant $\alpha$. This has no effect  
on the cars on the road and their fluctuations
because all removed cars are taken from the parked ones residing 
inside the garage. These surplus cars are dynamically inert.
This continues until the reservoir is depleted, 
the point at which the transition to the $N$ phase takes place.
The road density becomes equal to the total car density $\rho_r=\rho_o$.
From the $C$ (input-limited) phase perspective,  
this happens at $\rho_o=\alpha$, see Fig.~\ref{newPD}, because 
$\rho_r=\alpha$ inside the $C$ phase~\cite{Derrida,Krug,Popkov-S}.
From the MC (road-limited) phase perspective, 
it occurs at $\rho_o=\frac{1}{2}$ 
because $\rho_r=\frac{1}{2}$ inside MC.
Therefore, the phase boundaries into the $N$ phase are located
at $\rho_o=\alpha$ and $\rho_o=\frac{1}{2}$.

\begin{figure}
\centerline{\epsfxsize=6.5cm \epsfbox{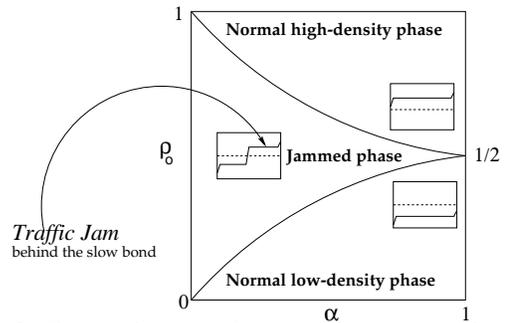}}
\caption
{Phase diagram of the Janowsky-Lebowitz model, an
AEP with periodic boundary conditions
and a slow bond. The corresponding density profiles 
are schematically shown in the insets.}
\label{pbcPD}
\end{figure}

The next issue is how many types of normal phases exist 
in the lower right side of the phase diagram. In a $N$ phase, 
the garage acts like a localized impurity (a blocking-type site), 
and it contains typically only a few cars.
Suppose we put a cap $N_m$ on the occupation of the garage,
$N_p=0,1,...,N_m$. This cannot affect the properties of the
model in the $N$ phase, except very close to the transition into the
condensate phases (where the number of parked cars $N_p$
and the fluctuations in $N_p$ diverge).
Janowsky and Lebowitz~\cite{J-Lebowitz} (JL)
studied such a capped version of the AEP model,
a closed loop road with periodic boundary conditions,
without a garage, but with one special bond 
where the hopping probability is reduced from 1 to $\alpha$. 
This is equivalent to setting the occupation limit of the garage 
to $N_m=1$. Their phase diagram has the same control parameters as ours.
It has particle-hole symmetry with respect to $\rho_o=\frac{1}{2}$,
see Fig.~\ref{pbcPD}.
The normal phase in the lower right hand corner is 
similar to our model. The second normal phase 
in the upper right corner is equivalent to it by particle-hole symmetry, 
and the intermediate area is a coexistence region, the jammed phase.
In the low-density $N$ phase, the cars are uniformly distributed,
but at $\rho_o=\alpha/(1+\alpha)$ a traffic-jam-type shock wave develops
between low- and high-density $N$-type phase regions.

This jammed phase is reminiscent of the condensate phase
(macroscopic occupation of the parking garage) in our model.
A shock wave does not appear in our model because the garage
absorbs the queue. It is also clear that if we would interpolate
between the JL model and our model by slowly increasing $N_m=1\to\infty$,
the queuing transition would be delayed and would be shifting smoothly,
with our $N$-$C$ phase boundary as the limiting case.
JL focused on the finite-size-scaling (FSS) properties
of the fluctuations in the position of the tail of the shock wave,
$\Delta l \sim L^{\gamma}$, and found $\gamma=\frac{1}{2}$ 
(except for $\rho_o=\frac{1}{2}$ where a cancellation of the
leading time of flight-type fluctuations gives rise to
$\gamma=\frac{1}{3}$)~\cite{J-Lebowitz}.
Monitoring the location of the shock wave is
analogous to observing the parked car density
fluctuations in our model inside the $C$ and MC phases.
JL did not address the scaling properties at the
queuing transition itself, but their transition is definitely
second order (the location of the shock wave moves continuously
with the total density away from the blockage point).

\section{KPZ Growth}
\label{KPZ}

The AEP can be interpreted as a lattice representation
of the Burgers equation with a discretized velocity field, both in 
location and in its values,
$n(x)=0,1$.
The latter is equivalent to the
so-called body-centered solid-on-solid growth
model~\cite{RBDetc,MdN}, a lattice realization of
KPZ-type surface growth.
Each occupied site represents a down step along 
the one-dimentional (1D) surface and
every empty site an up step. The surface heights, $h(x+\frac{1}{2})$,
are limited to even/odd values at even/odd lattice sites,
$h(x+\frac{1}{2})- h(x-\frac{1}{2})=1-2n(x)$.
The surface looks like the alternatively
stacked bricks of a masonary wall, but rotated over 90 deg.
Each car jump to the right represents 
the deposition of a new vertical 1x2 brick.
The average car density on the road represents the average slope
of the surface. At half density the surface is not tilted.

Periodic boundary conditions are the most common in theoretical studies of
surface growth, and this brick deposition version of KPZ growth
is exactly soluble by the Bethe ansatz~\cite{G-Spohn}.
It is a special line in the general phase diagram of the 1D
well-known {\em XXZ} spin-$\frac{1}{2}$ quantum spin chain Hamiltonian,
and also in the so-called six-vertex model of 2D equilibrium critical
phenomena~\cite{XXZ-6vertex}.

From the surface growth perspective, 
the open road setup with reservoirs represents 
an open ended 1D interface with modified particle deposition
probabilities $\alpha$ and $\beta$
at its end points. In the MC phase, where the
car density is locked to $\rho_o=\frac{1}{2}$, the
crystal surface maintains a net zero tilt.
For $\alpha<\frac{1}{2}$ and/or $\beta<\frac{1}{2}$,
the reduced growth probabilities at the edge create a
nonzero globally tilted surface. Moreover, along the line
$\alpha=\beta<\frac{1}{2}$, two tilts coexist. Crossing that line amounts
to undergoing a first-order phase transition with different tilt angles.
The lock-in transition from the $C$ phase into the MC phase represents
a second-order dynamic facetting transition.

Our parking garage setup translates into a surface growth dynamics
realization with periodic boundary conditions and a localized defect.
The garage could represent something like a stacking defect, where
the step height is unlimited (and somehow energetically cost free).
The cliff height at the stacking fault can be microscopic (the $N$ phase)
or macroscopic (the $C$ and MC phases). This depends on the modified
growth probability $\alpha$ near the defect (on top of the cliff,
somewhat similar to a Schwoebel energy barrier)
and also depends on the global net tilt angle (the total car density).
In the MC ($C$) phase, the average slope of the surface,
excluding the cliff, is zero (nonzero).

\section{Correlations and Group Velocity}
\label{groupvelocity}

It is well known and easy to show that the
stationary state of KPZ growth with PBC,
the closed loop road without obstacles where $\alpha=1$ and $n(N_s)=0,1$,
is completely random without any correlations
whatsoever between finding a car at sites $i$ and $j$ for any $i\neq j$
including nearest neighbor sites.
In the surface representation, this means that
up and down steps are placed at random~\cite{MdN}
and that the width of the interface $W\sim N_s^{\chi}$
scales with the random walk exponent $\chi=\frac{1}{2}$.

The stationary state remains uncorrelated in our model as well (as 
shown below) except near the garage. Those correlations are governed by 
the dynamic scaling exponent and the group velocity with which 
fluctuations travel along the road.

In 1D KPZ-type surface growth, all characteristic times scale as
$t\sim N_s^z$ with $z=3/2$. In car language, 
this means that local car density fluctuations
with spatial width $l$ broaden in time as $l\sim t^{1/z}$.
The value of the dynamic exponent $z$
follows from the  identity $z+\chi=2$ implied
by Galilean invariance of the Burgers equation~\cite{G-Spohn,K-P-Zhang},
together with the disordered $\chi=\frac{1}{2}$ nature of the stationary state.
The exact (Bethe ansatz) solution of this specific model confirms this
general result.

Fluctuations travel with a group velocity $v_g=1-2\rho_r$ 
to the right (towards the garage) along the road.
From the KPZ surface growth perspective,
the group velocity represents the average slope of the KPZ growing 
surface, and its fluctuations grow perpendicular to the surface 
orientation. From the AEP formulation perspective, the value of $v_g$ is 
set by the average stationary state current $j=\rho_r(1-\rho_r)$
(easy to derive since the stationary state is Gaussian)
and the definition of the group velocity is
\begin{equation}
v_g=\frac{\partial}{\partial \rho_r}j(\rho_r)=(1-2\rho_r).
\end{equation}
This $v_g$ played an important role, e.g., in the AEP study by
Majumdar and Barma~\cite{M-Barma} of tagged-particle diffusion, and
in describing phase transitions between
steady states in the open road setup~\cite{K-Schuetz-K-S}.

\section{Density profiles in the various phases}
\label{densityprofile}

The on-the-road car density profiles in the $C$ and MC phases
are known exactly from the open boundary model studies,
in particular the one by Derrida {\em et al.}~\cite{Derrida}.
These profiles play an important role in our later discussions
and it is useful to review how the group velocity enters in
qualitative explanations of these exact results.

In the road-limited $MC$ phase at
$\alpha>\frac{1}{2}$, the density tail has a power-law shape,
\begin{equation}
\rho(x)\simeq \rho_r + A_e x^{-\nu},
\label{density}
\end{equation}
at both edges $e$=$L$(left),$R$(right),
with $x$ the distance from the edge of the road 
(garage entrance in our model)
and $\rho_r=\frac{1}{2}$ the on-the-road car density.
In contrast, in the input-limited $C$ phase at $\alpha<\frac{1}{2}$,
the car density is constant, $\rho(x)=\rho_r$, at the beginning of the 
road (garage exit), and has an exponential tail 
at the end of road (garage entrance),
\begin{equation}
\rho(N_s-x)\simeq \rho_r+A x^{-3/2} e^{-x/\xi},
\label{density-exp}
\end{equation}
with the on-the-road car density $\rho_r=\alpha$ 
and the correlation length $\xi =-1/\ln[4\alpha(1-\alpha)]$. 

Power laws with exponent $\nu=\frac{1}{2}$ arise
naturally in this problem because of its critical fluctuations.
The bulk properties of the KPZ stationary state are invariant
under a rescaling of all lengths as $x^\prime =b x$, all times
as $t^\prime =b^z t$, and the surface heights as $h^\prime= b^\chi h$.
The car density scales, therefore, naively as
$\rho=\partial h/\partial x \sim b^{\chi-1}$.
This means that power-law tails in the density distribution near 
the edge of the road,
with exponent $\chi-1=-\frac{1}{2}$, like in Eq.~(\ref{density}),
are to be expected.

On the other hand,
the disordered nature of the stationary state
suggests exponential profiles, like in Eq.~(\ref{density-exp}).
Indeed, in the bulk, 
the stationary state density-density correlation function
\begin{equation}
g(r) = \left< \rho(x+r)\rho(x) \right> -
\left< \rho(x+r)\right> \left<\rho(x) \right>
\end{equation}
does not decay as a $r^{-1}$ power law as may be
suggested by the above scaling argument, but instead
decays exponentially with a very short correlation length,
that is zero in this specific model 
since the cars on the road are totally uncorrelated in the stationary state.

To make sense of Eqs.~(\ref{density}) and (\ref{density-exp}),
it is important to realize that the density profile
near the edge of the road incorporates temporal correlations,
and also to appreciate the role of the group velocity.

The power-law profile in the MC phase is the result of
correlations with cars that reached the edge of the road 
and moved into the garage at earlier times.
Such correlations spread in time over a spatial distance $l\sim t^{1/z}$.
The group velocity is zero in the MC phase and therefore the cars
at a distance $l$ from the edge of the road are (power-law) correlated 
with those that entered the garage at times earlier than $\Delta t= l^z$. 
They have no knowledge about cars arriving at the garage more recently.
This explains the power-law tail in the density distribution.

In the $C$ phase, the same correlation spreading takes place 
within a moving frame of reference with nonzero group velocity.
Information reaches the edge of the road (garage entrance) at 
a rate $v_g$ (linear in time) faster than it can spread backwards
as $l\sim t^{1/z}$ (since $1/z=2/3<1$).
Memory has no opportunity to develop near the edge of the road,
and the density profile adjusts itself at the road end 
as if the cars on the road are completely uncorrelated.
The exponential tail reflects a suction-type effect;
the excluded volume limitation on the car mobility does not act 
on the cars departing from the road.

Similarly, in the MC phase, power-law correlations with
earlier cars that escaped from the garage and entered the road 
give rise to a power-law tail in the density profile 
at the beginning of the road (near the exit of the garage), 
while in the $C$ phase, 
this information travels faster away from the road start than it can
spread backwards. So, new cars entering the road do so completely
uncorrelated, resulting in no road-start tail  
in the profile whatsoever. This confirms why the MC phase is
road limited and the $C$ phase is input limited.
In the latter, the car supply at the garage controls 
the density near the edge of the road and 
also everywhere else on the road.

\begin{figure}
\centerline
{\epsfxsize=6cm \epsfbox{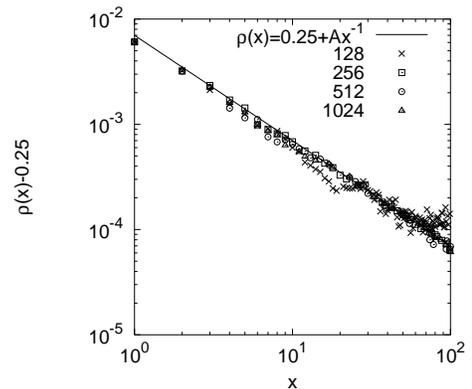}}
\caption
{Double logarithmic plots of the density profiles 
in the $N$ phase ($\rho_o=0.25$ and $\alpha=1$),
showing a $1/x$ tail at the beginning of the road (garage exit).}
\label{Nnu}
\end{figure}

In the $N$ phase, we find numerically, from Monte Carlo simulations,
an exponential tail at the road end and
an $1/x$ tail at the road start as shown in Fig.~\ref{Nnu}.
The group velocity is also nonzero in the $N$ phase, which explains
the exponential exit tail as follows. Just like in the $C$ phase, the 
garage is invisible to the incoming of cars from the road; fluctuations 
arrive at the garage faster (at constant velocity) than they can spread 
backwards (with $\Delta l \sim t^{1/z}$).
The fact that in the $N$ phase only a few cars reside inside the 
garage is invisible to the cars entering the garage.

An $1/x$ tail at the beginning of the road 
is what one expects from the deterministic part of the Burgers equation.
The solution of the deterministic Burgers equation,
\begin{equation}
\frac{\partial v }{\partial t} + \lambda v \frac{\partial v}{\partial
x}= \nu \frac{\partial^2v}{\partial x^2}
\label{Burgers}
\end{equation}
with the velocity $v$ pinned at a specific nonzero value
at site $x=0$, is readily seen (e.g., in the Hopf transformed
formulation) to be of the generic form
$v(x)\sim 1/(1+ax)$, i.e., having the $1/x$ power-law shape.

In the $C$ phase, the system selects the value $a=0$ for the constant,
and in the $N$ phase, $a\neq 0$. In both cases, 
the group velocity $v_g>0$ carries away KPZ fluctuations
faster than they can spread, such that the KPZ noise term can be
ignored at the road start (stationary frame of reference).
The only noise that remains is that from the random process by which
cars are being taken out of the garage and put on the road.
In the $C$ phase, 
the supply of cars is bottomless ($\rho_p>0$), such that $a=0$
(the entire on the on-the-road bulk car density is ruled by the garage),
while, in the $N$ phase, the supply of cars is limited and 
the garage does not overwhelm the road, such that $a\neq 0$.

\section{Density-density correlations}
\label{correlations}

It is useful to discuss how the group velocity affects 
the car-car correlation function,
\begin{equation}
g(r,\tau) = \left< \rho(x+r)_{t+\tau}\rho(x)_t \right> -
\left< \rho(x+r)_{t+\tau}\right> \left<\rho(x)_t \right>.
\label{density-density}
\end{equation}
According to scaling theory, it obeys the form
\begin{eqnarray}
g(r,\tau)&=& b^{2(\chi-1)} g(b^{-1}r,b^{-z}\tau)\nonumber\\ 
         &=&\tau^{\frac{2(\chi-1)}{z}} F(\frac{r}{\tau^{1/z}}).
\label {g-scaling}
\end{eqnarray}
In the limit $\phi=r/\tau^{1/z}\rightarrow 0$,
the scaling function $F(\phi)$ must approach a constant
since the autocorrelation decays as a power law,
\begin{equation}
g(0,\tau)\sim \tau^{\frac{2(\chi-1)}{z}}\sim \tau^{-2/3},
\end{equation}
with $\chi=\frac{1}{2}$.
In the opposite limit, of large $\phi$,
the scaling function must decay exponentially
because $g(r,0)=0$ due to the random nature of the stationary state
(or, more  generically, it decays  exponentially with 
some correlation range $\phi_o$ of the same order as the interaction
range between cars).
The above scaling relation suggests the form
\begin{equation}
F(\phi) \sim \phi^{2(\chi-1)} e^{-\phi/\phi_o} 
\end{equation}
in the limit $\phi=r/\tau^{1/z}\rightarrow\infty$,
such that
\begin{equation}
g(r,\tau)\sim r^{2(\chi-1)} 
e^{-r/\phi_o\tau^{1/z}}\sim \frac{1}{r} e^{-r/l}
\end{equation}
with $\chi=1/2$ and $l=\phi_o\tau^{1/z}$. 
The length $l$ in the exponential defines the ``correlation cone" 
that we already mentioned above. Within the cone 
the correlations decay as a power law, and outside it 
the cars are uncorrelated (as in the stationary state).
The above discussion ignores finite-size effects.
For open road or garage type boundary conditions,
the correlator explicitly depends on the initial position $x_0$
of the car at time $t_0$ and also on the road length $N_s$,
\begin{equation}
g(r,\tau ;x_0,N_s)= b^{2(\chi-1)} 
g(b^{-1}r,b^{-z}\tau ; b^{-1} x_0, b^{-1} N_s).
\label {g-fss-scaling}
\end{equation}

\begin{figure}
\centerline
{\epsfxsize=6cm \epsfbox{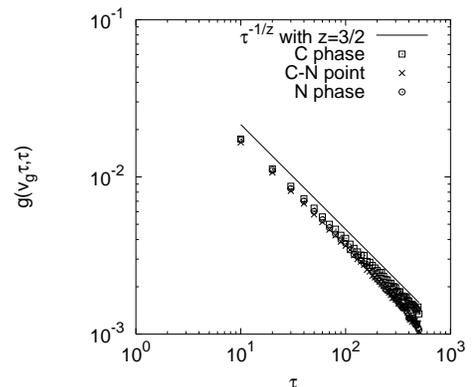}}
\caption
{Double logarithmic plots of
the density-density correlation function
$g(r,\tau)$ in the moving frame of reference, 
i.e., with $r=v_g\tau$ and $v_g=(1-2\rho_r)$. 
The $C$ phase data are obtained at $\rho_o=0.75$ and $\alpha=0.25$, 
the $C$-$N$ point data at $\rho_o$=$\alpha$=0.25, 
and the $N$ phase data at $\rho_o=0.25$ and $\alpha=1$.}
\label{plot_g}
\end{figure}

Finite-size effects set in at times
when the correlation cones hit the road edges,
$\tau\sim (x_0)^{z}$ or $\tau\sim ( N_s-x_0)^{z}$.
At times longer than $t_0\sim N_s^{z}$, we therefore expect
that the car-car correlator decays exponentially, e.g., as
\begin{equation}
g\left( 0,\tau ;\frac{1}{2}N_s,N_s \right) \sim N_s^{2(\chi-1)} 
e^{-a(\tau/N_s^{1/z})}.
\label{g-fss-exp}
\end{equation}
The above analysis applies to the MC phase, 
where the drift (group) velocity is zero. In the $C$ phase 
(and also the $N$ phase), we need to switch to the moving frame 
of reference by replacing $r\rightarrow r+v_g\tau$.
This has some peculiar consequences. For example, 
the autocorrelation function $g(0,\tau)$ decays exponentially in time
(even at short times). The correlation cone $l$ is slanted in the 
direction of the flow (the correlations move with the flow)
and $l$ widens slower than linear in time, 
such that the $r=0$ line in the world sheet lies 
outside the correlation cone.
The KPZ-type correlations are somewhat hidden. 
In order to expose them, one needs to plot them 
in some special manner, e.g., $g(v_g\tau,\tau)$
(the autocorrelator in the moving frame of reference),
as illustrated in Fig.~\ref{plot_g}.

\section{Scaling at and near the Condensation Transitions}
\label{fsscaling}

Let us turn now to the scaling properties 
of the two condensate phase transitions from the $C$ and MC phases 
into the $N$ phase. The stationary state value of the parked car density 
$\rho_p=N_p/N_s$ acts as the order parameter.
We expect it to obey the FSS relation
\begin{equation}
\label{scaling}
\rho_p (\epsilon, N_s ^{-1})=
b^{-x_p}\rho_p (b^{y_\epsilon}\epsilon, bN_s^{-1})
\end{equation}
near the two condensation transition lines
with a scaling factor $b$, and
$\epsilon=(\alpha-\alpha^*)~ {\rm or}~ \epsilon=(\rho^*_o-\rho_o)$,
a measure of the distance from the transition.

\begin{figure}
\centerline
{\epsfxsize=4.cm \epsfbox{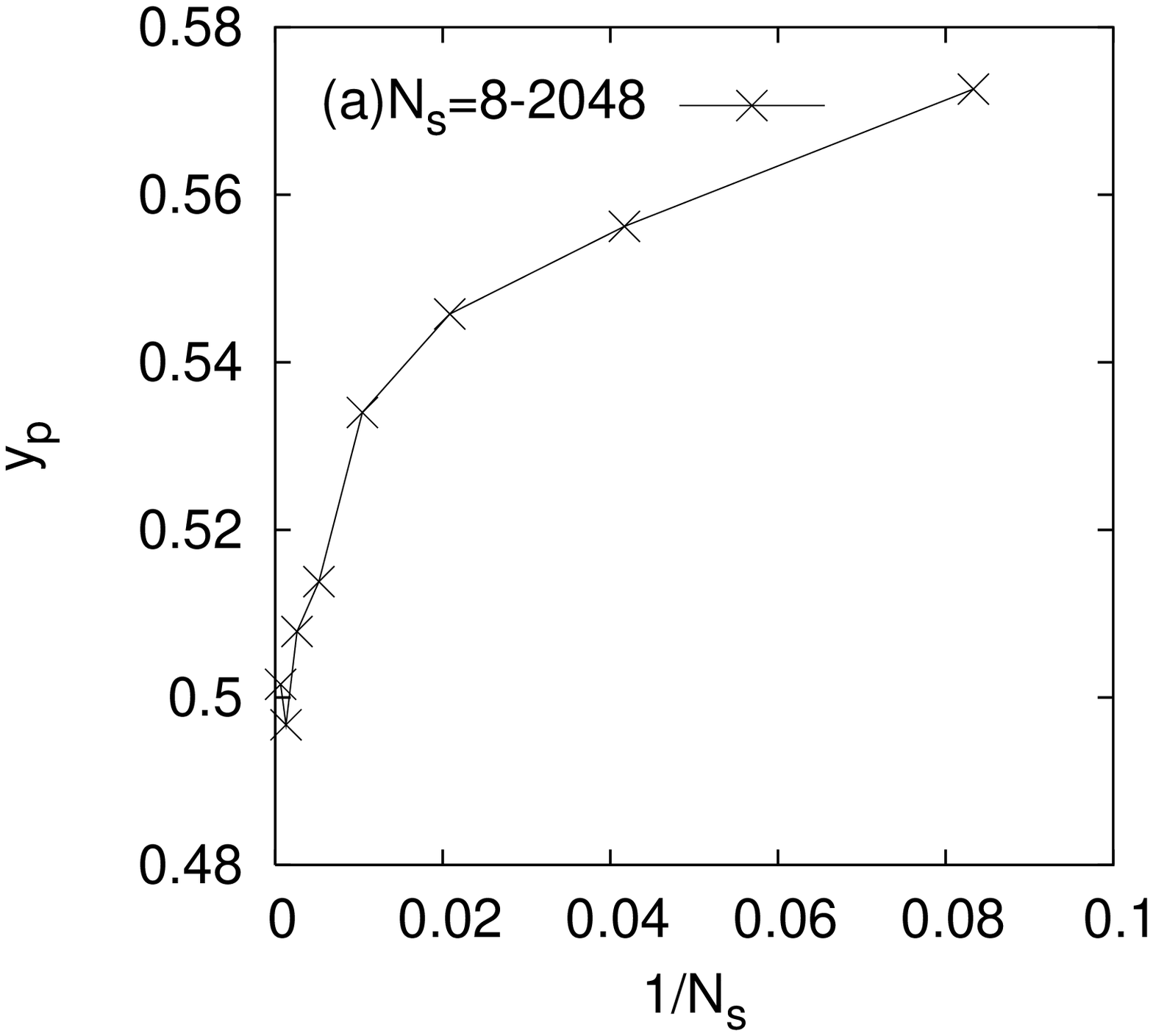} 
\epsfxsize=4.2cm \epsfbox{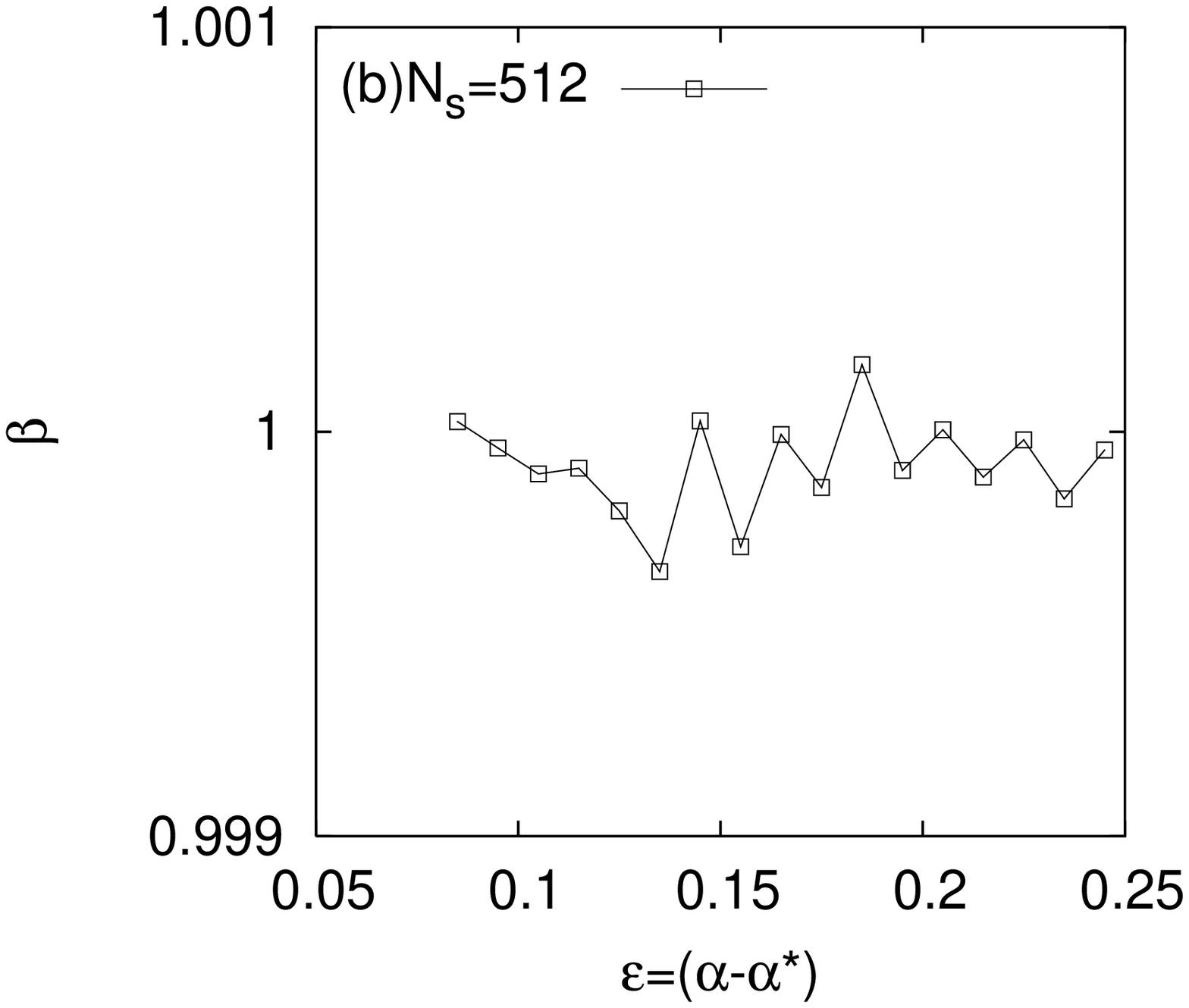}}
\caption
{The parked car (order parameter) exponents
(a)~$y_p=0.505(5)$ and  (b)~$\beta=1.000(1)$, defined as $N_p\sim 
N_s^{y_p}$ and $\rho_p\sim (\alpha-\alpha^*)^{\beta}$,
at $\rho_o$=$\alpha$=0.25 of the $C$-$N$ phase boundary.}
\label{y_CN}
\end{figure}

\begin{figure}
\centerline
{\epsfxsize=4cm \epsfbox{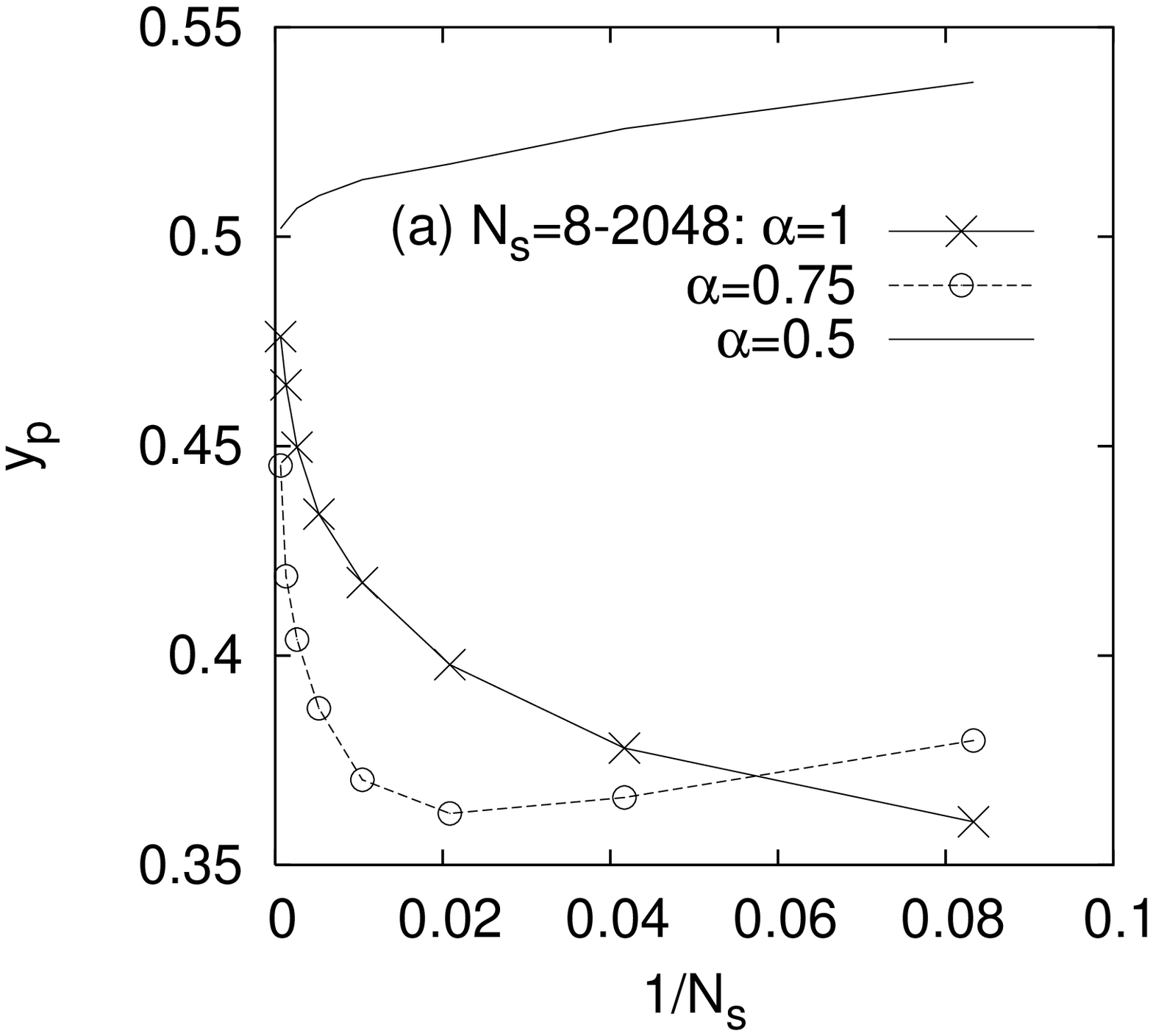} 
\epsfxsize=4.2cm \epsfbox{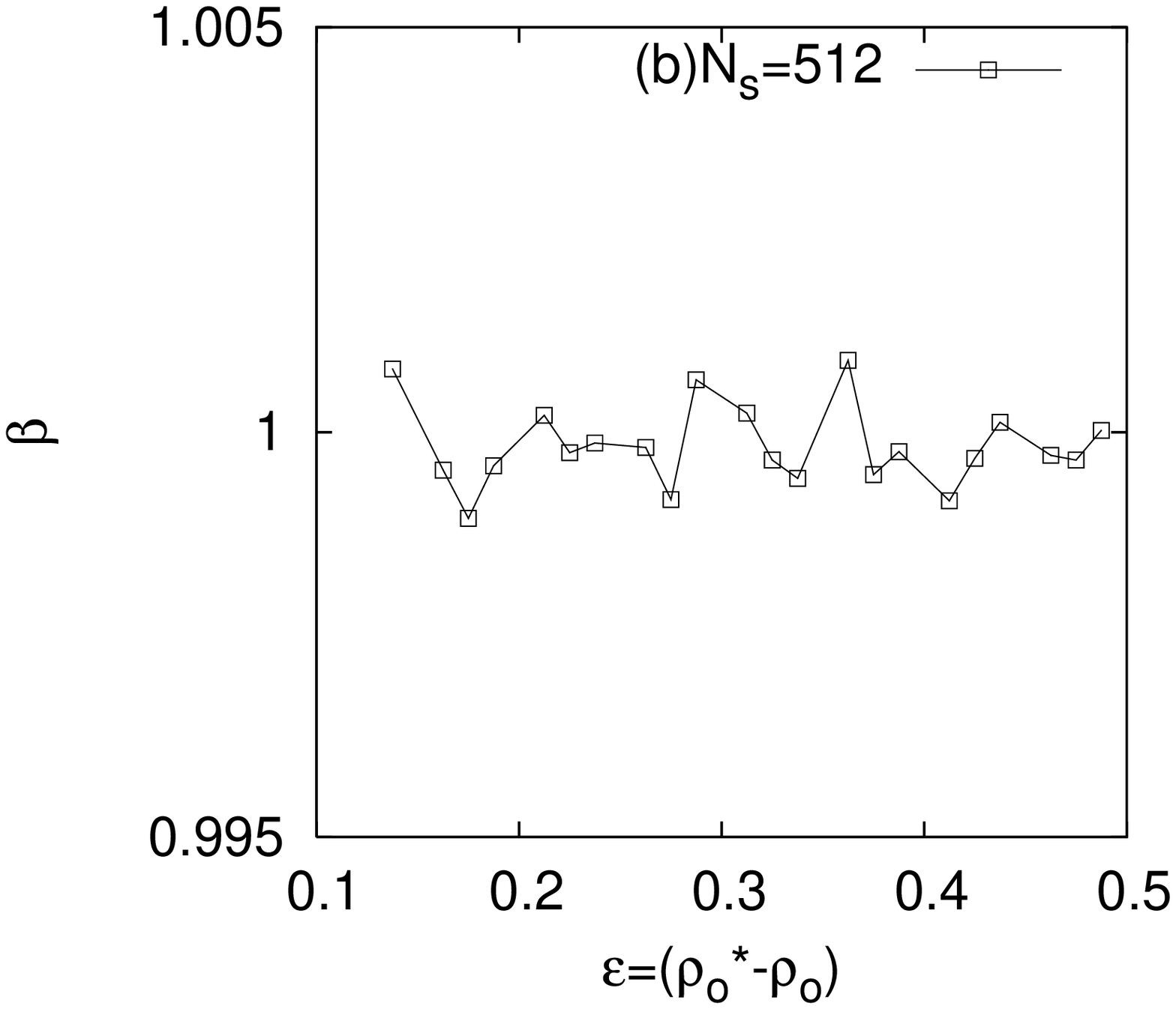}}
\caption
{The parked car (order parameter) 
exponents (a)~$y_p=0.495(5)$ and (b)~$\beta=0.997(3)$,
at $\rho_o=1/2$ and $\alpha=1$ of the $N$-MC phase boundary. 
For other two curves in (a), taken along 
the MC-$N$ phase boundary at $\alpha\ge 1/2$,
the exponent $y_p$ retains the same value, but is subject to 
large/small corrections to finite-size scaling originating from 2/3 
power-law density profile at the beginning of the road 
(garage exit).}
\label{y_NMC}
\end{figure}

In the $N$ phase, the density of parked cars is zero, $\rho_p=0$.
When approached from the $C$ and MC sides,
$\rho_p$  goes to zero as $\rho_p\sim 
|\epsilon|^{\beta}$, where $\beta={x_p/y_\epsilon}$.
The removal of cars from the road is accommodated by taking them from the
passive inert ones residing inside the garage. This implies that
$\rho_p$ must vanishes linearly at the transition, and 
that $x_p=y_\epsilon$. Our numerical simulations confirm that 
$\rho_p\sim |\epsilon|$, as $\beta=1$ 
in Figs.~\ref{y_CN} and \ref{y_NMC}.

The total number of parked cars ($N_p$)  scales as
\begin{equation}
\label{Nscaling}
N_p (\epsilon, N_s ^{-1})=
b^{y_p} N_p (b^{y_\epsilon}\epsilon, bN_s^{-1})
\end{equation}
with exponent $y_p=1-x_p$ 
according to Eq.~(\ref{Nscaling}) and $N_p=\rho_pN_s$.
In the $N$ phase, the density $\rho_p$ remains zero, while
the total number of parked cars ($N_p$) diverges towards the transition 
as $N_p\sim |\epsilon|^{(1-x_p)/y_\epsilon}$.
We find numerically that this power law is linear as well, 
$N_p\sim |\epsilon|$.
Combined with the linear scaling of the car density
from the $C$ or MC side (implying $y_p=x_p$),
this yields $y_p=y_{\epsilon}=\frac{1}{2}$.
The exponent $y_p$ determines the FSS behavior
of the total number of parked cars at the transition point itself,
$N_p\sim N_s^{y_p}$ (and also that the density of parked cars vanishes 
as $\rho_p\sim N_s^{y_p-1}$). 

Figure~\ref{y_CN} shows the numerical 
results of $y_p$ and $\beta$ at point $\alpha=0.25$
of the $C$-$N$ phase boundary and Fig.~\ref{y_NMC} at 
$\alpha=1$ (and 0.75) of the MC-$N$ phase boundary. 
Notice that the FSS corrections in the exponent $y_p$ are much stronger 
at the MC-$N$ transition than at the $C$-$N$ transition.

\begin{figure}
\centerline
{\epsfxsize=3.7cm \epsfbox{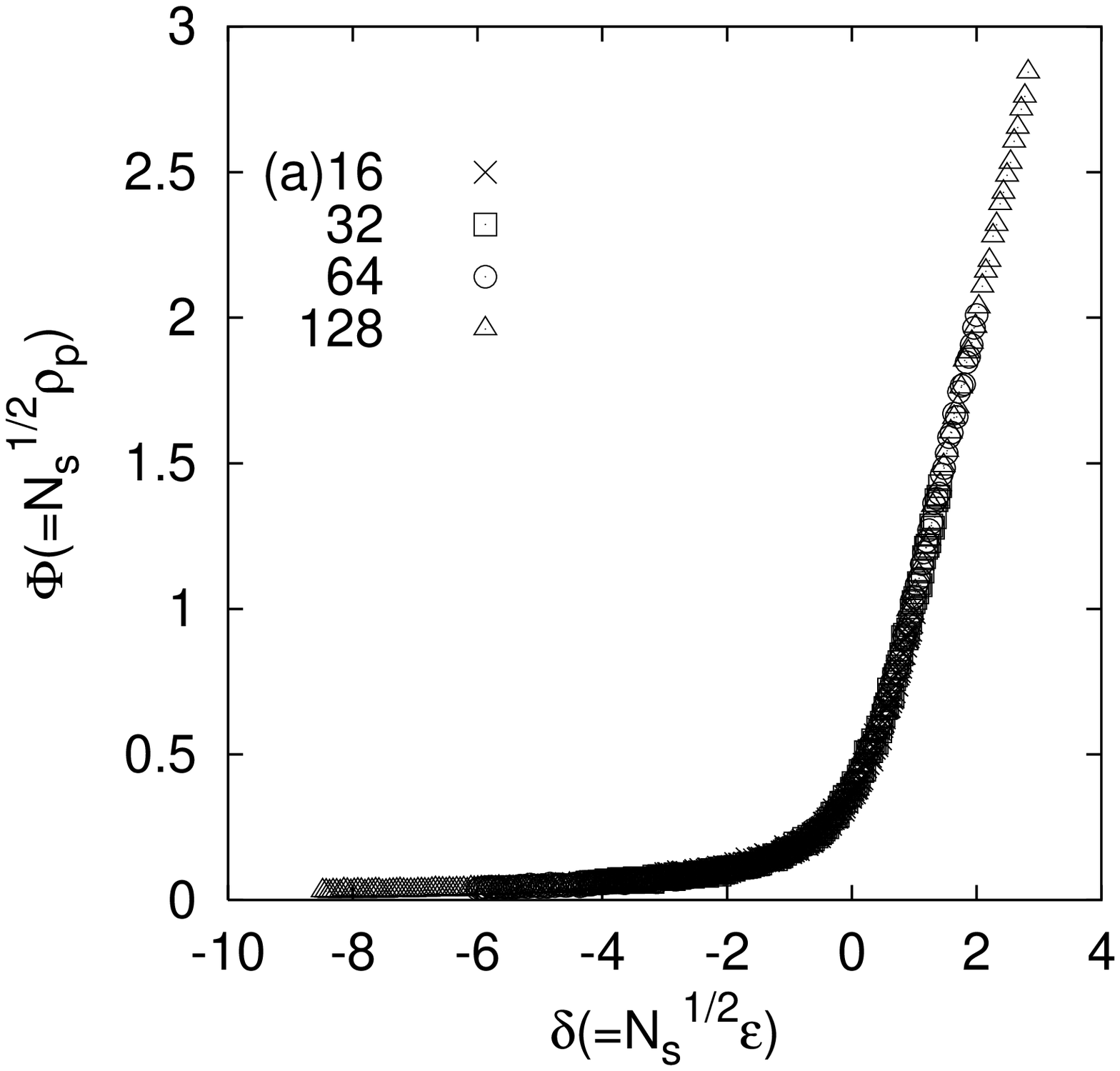} 
\hspace*{0.75cm}\epsfxsize=3.45cm \epsfbox{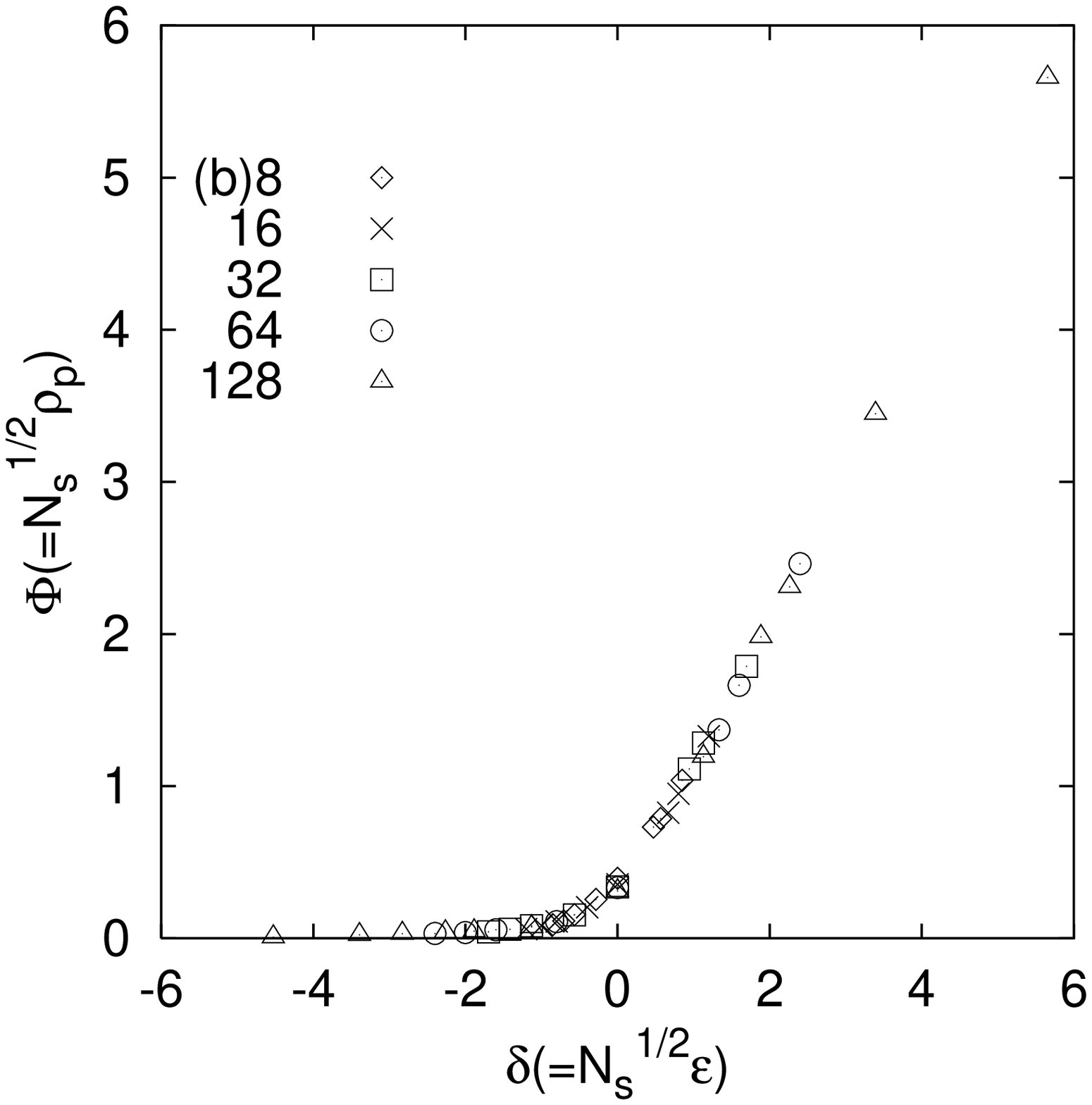}}
\caption
{Scaling function $\Phi(\xi)$ defined in Eq.~(\ref{FSS})
at the same transition points as in the previous two figures; 
(a)~$\rho_o=\alpha=0.25$~($C$-$N$) and 
(b)~$\rho_o=1/2,~\alpha=1$~(MC-$N$).
The numerical data collapse very well for $x_p=y_\epsilon=1/2$, 
as suggested in Eq.~(\ref{FSS}). The scaling function 
$\Phi(\xi)\to 0$ for $\xi\ll 0$ and $\Phi(\xi)\to \xi$ for $\xi\gg 0$.} 
\label{collapse}
\end{figure}

The validity of the scaling relations is further illustrated
by plotting the scaling function $\Phi(\xi)$ in the following form
\begin{equation}
\label{FSS}
\rho_p (\epsilon, N_s ^{-1})=
N_s^{-x_p} \Phi(N_s^{y_{\epsilon}}\epsilon),
\end{equation}
and the associated data collapse while moving through
the transition points
at constant $\rho_o$ and constant $\alpha$, respectively.
The curves in Fig.~\ref{collapse} collapse very well.

The fluctuations and the FSS corrections to the number of parked
cars are a mirror of the density-density correlations and
the density profile of cars on the road. They also reflect how the latter 
builds up as a function of the length of the road.
At the transition points, the bulk value of the on-the-road density 
$\rho_r=N_r/N_s$ is exactly equal to the total number of cars in the 
system divided by the road length, $\rho_r=N_c/N_s$,
such that there would be no need for any car to remain inside the garage.

We find numerically that at the $C$-$N$ transition point, 
the density profile retains the same structure as inside the $C$ phase;
with no tail at the beginning of the road (garage exit) 
and an exponential tail at the end of the road
(garage entrance). Such profiles cannot account for
the $y_p=\frac{1}{2}$ FSS divergence in the number of parked cars.
The scaling behavior $N_p\sim N_s^{y_p}$ must, therefore, reflect
directly the corrections to FSS in the bulk density of cars on the road.
The value $x_p=\frac{1}{2}$ naturally arises because
at the transition point, $\rho_p\sim \rho_r$,
and $\rho_r=\partial h/ \partial x$ scales as 
$L^{\chi-1}\sim L^{-x_p}$, where $L$ 
corresponds to a given length of the road $N_s$.

A more intuitive explanation follows again from the
nonzero group velocity at the $C$-$N$ phase boundary.
As mentioned above in the discussion of the density
distribution tail in the $C$ phase, the events by which
cars enter the road from the garage are completely uncorrelated
due to the slanting of the correlation cones
(the correlations move with the flow towards the garage and spread 
slower than linearly, only as $l\sim t^{1/z}$, such that communications 
with later events at the beginning of the road are impossible).
So the entry events to the road from the garage behave 
like uncorrelated random noise,
and the fluctuations in the number of cars scale, 
therefore, as the square root of time. The time in question is 
proportional to $N_s$ because the fluctuations 
(created at the garage exit) move along the road with 
velocity $v_g$, and are wiped out after they return to the garage. 
From this, it follows that the fluctuations in the number of parked cars 
scales as $\sqrt N_s$. Moreover, $N_p$ cannot be negative, which means 
that the fluctuations sample the bottom of the garage 
and that the transition from the $C$ to $N$ phase, therefore, 
takes place when the garage contains $N_p\sim \sqrt N_s$ cars.

In summary, the scaling at the $C$-$N$ phase boundary is governed by the
bulk fluctuations in the-on-the-road density, which is ruled by
the nonzero group velocity of KPZ fluctuations, and
this leads directly to random noise like $N_p\sim \sqrt N_s$, 
corrections to FSS.

\begin{figure}
\centerline
{\epsfxsize=6cm \epsfbox{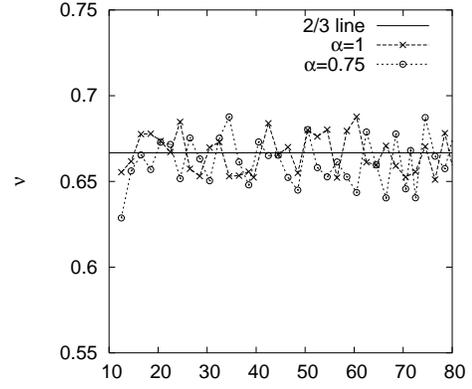}}
\caption
{Effective (finite-size-scaling type) value for the exponent 
$\nu$ of the density profiles at the beginning of the road: 
$\nu=0.66(1)$ at $\rho_o=1/2$ and $\alpha=1$ (or 0.75). 
At the MC-$N$ phase boundary, the density profiles decay as 
a 2/3 power law, instead of the 1/2 power found inside the MC phase.}
\label{NMCnu}
\end{figure}

The FSS corrections to scaling in exponent $x_p$ ($y_p$) 
at the MC-$N$ phase boundary are much more complex.
The (bulk) group velocity is zero, and power-law density profiles
are realized at both edges of the road.
At the end of the road, the density profile 
follows a critical exponent $\nu=\frac{1}{2}$, 
the same power as that inside the MC phase discussed in 
Sec.~\ref{densityprofile}. However, at the road start 
(the exit of the garage),
the power-law exponent changes from $\nu=\frac{1}{2}$ inside 
the MC phase to $\nu=\frac{2}{3}$ at the MC-$N$ transition.
This is shown in Fig.~\ref{NMCnu} for $\alpha=1$ and
$\rho_o=\frac{1}{2}$, where the effect is the strongest.

The $\frac{2}{3}$ power law does not change the $N_s^{1/2}$ FSS behavior 
of the number of parked cars. It is responsible, however,
for strong corrections to FSS, as clearly visible in Fig.~\ref{y_NMC}(a).
The $\nu=\frac{2}{3}$ power-law profile contributes 
only a subdominant term to $\rho_p$ because it decays faster 
than the two $\nu=\frac{1}{2}$ contributions
(from the density profile at the end of the road and
from the KPZ-like bulk road density fluctuations).
We have not achieved yet a good understanding 
of this novel value, $\frac{2}{3}$, for the exponent of 
the density profile at the beginning of the road.
It obviously lies correctly in between the MC and $N$ values,
$\frac{1}{2}$ and $1$, respectively.
Moreover, its value, $\nu=\frac{2}{3}$, is likely linked to
the KPZ dynamic exponent $z=\frac{3}{2}$.
But it remains unclear how to glue the following pieces together.

At the MC-$N$ transition, the group velocity of the fluctuations, 
$v_g=1-2\rho_r$, is still zero (but only barely) 
since $\rho_r=\frac{1}{2}$. This means that time of flight aspects, 
which dominate the $C$ and $N$ phases, do not come into play. 
Near the garage, the tail in the density profile,
$\rho(x)=\frac{1}{2}+\Delta \rho(x)$, creates a local nonzero
group velocity $v_g(x)= -2\Delta \rho(x)$ 
pointing back into the garage
(instead of away from it as in the $C$ and $N$ phases).

Recall from Sec.~\ref{densityprofile} 
that the density profile in the MC phase has the same type 
of power-law profile $\Delta \rho(x)\sim x^{-\nu}$,
but with the ``KPZ" (power-counting) exponent value $\nu=\frac{1}{2}$.
In that case, fluctuations at the road start do not reach 
the bulk of the road: the leading edge of the information cone 
is stationary because the backward movement of its center of mass
$x_c\sim -t^{1/(1+\nu)}$ (implied by $dx_c/dt\sim x_c^{-\nu}$)
matches exactly the rate of its spreading, $x\sim t^{1/z}$.
So the $\nu=\frac{1}{2}$ density profile fully screens 
the garage from view in the bulk of the road.
Exactly the same screening of the information takes place
at the end of the road (garage entrance) 
since also there the MC phase density profile has exponent 
$\nu=\frac{1}{2}$ (but in an opposite forward moving $v_g$ sense 
and with a negative density profile amplitude).

The density profile exponent $\nu=\frac{2}{3}$ at the MC-$N$ 
transition does not fully screen the garage any more from observers 
located far away on the road. Total screening is not needed
because KPZ fluctuations start to tunnel through the garage 
since it has only a $\sqrt N_s$ occupation. It is yet unclear to us, 
however, how to deduce (in a convincing manner) the exponent 
$\nu=\frac{2}{3}$ from these considerations.

\section{parked car fluctuations}
\label{carfluctuations}

In this section, we explore the fluctuations 
in the parked car density and also car-car correlations on the road. 
Of particular interest is the onset of transmission of information 
through the garage at the phase transitions. In the two condensate phases, 
the garage acts as a car reservoir and as a sink of fluctuations, while 
in the normal phase it contains only a few cars and transmits 
fluctuations.

The temporal fluctuations in the total number of parked cars, 
$G(N_s,\tau)$,
measures also the fluctuations in the total number of cars on the road.
It is  therefore equal to the integrated car-car
correlator (defined in Sec.~\ref{correlations}),
\begin{eqnarray}
G(N_s,\tau)
&=& \sum_{x_1,x_0=1}^{N_s-1}[\left<\rho(x_0,t_0)\rho(x_1,t_1)\right>-\bar 
\rho(x_0)\bar \rho(x_1)] \nonumber\\
&=& \sum_{x_0=1}^{N_s-1} ~~\sum_{r=-x_0+1}^{N_s-x_0-1}g(r,\tau;x_0,N_s)
\label{GG}
\end{eqnarray}
with $\tau=t_1-t_0$ and $r=x_1-x_0$.
The summations run over all road sites $x_0$ and distances $r$
that fit on the road.
Direct integration of the scaling relation,
Eq.~(\ref{g-fss-scaling}), yields that (in the MC phase)
$G$ obeys the scaling form
\begin{equation}
G(N_s,\tau)= b^{2\chi}~ G(b^{-1}N_s,b^{-z}\tau)
=N_s^{2\chi}~ {\cal F}(\tau/N_s^z).
\label {GG-scaling}
\end{equation}
In the KPZ representation, $G$ is the global slope-slope autocorrelator,
and at $\tau=0$, reduces to the conventional definition of interface width
(second moment of the height distribution).
We will now discuss how  $G$ behaves in the various phases
and at the phase transition points.

\subsection{Inside the MC phase}

The following intuitive discussion tells us 
how the scaling function ${\cal F}$ behaves in the MC phase.
Each $g(r,\tau)$ has a correlation cone of size $l\sim \tau^{1/z}$.
The cars within this cone are correlated with the one 
at site $x_0$ at time $t_0$
as $\tau^{2(\chi-1)/z}$. The integration over $x_0$ and $r$ 
in Eq.~(\ref{GG}) yields
\begin{equation}
G\sim N_s \times l \times \tau^{2(\chi-1)/z} \sim N_s \tau^{(2\chi-1)/z},
\label{timezero}
\end{equation}
where the correlation cones $l\sim\tau^{1/z}$ are assumed to be
small with respect to the road size $N_s$.
The exponent is equal to $\chi=\frac{1}{2}$, such that  $G\sim N_s$, and that
${\cal F}$ approaches
a constant in the limit $\tau/N_s^z \rightarrow 0$.

\begin{figure}
\centerline
{\epsfxsize=4cm \epsfbox{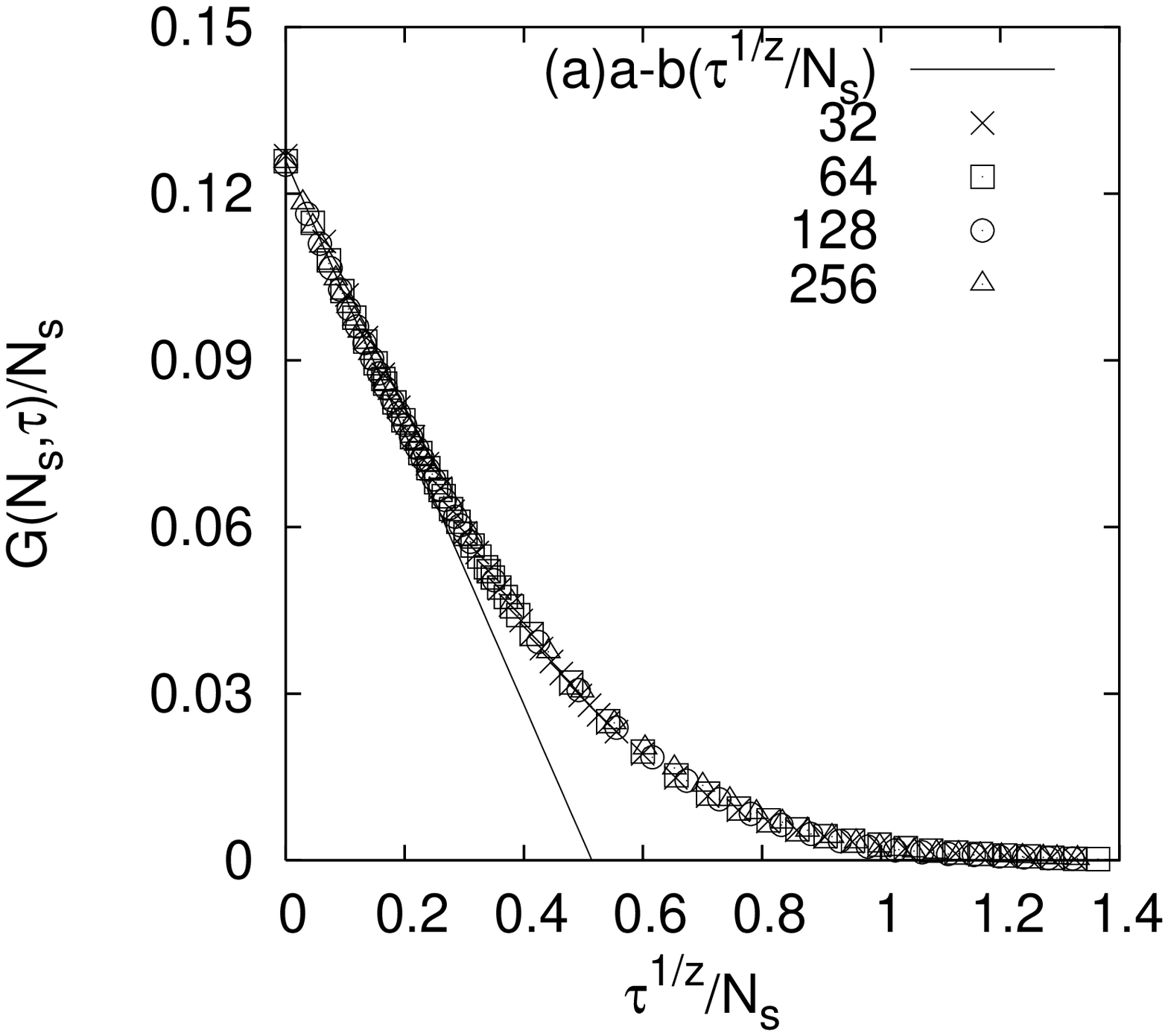} 
\epsfxsize=4.25cm \epsfbox{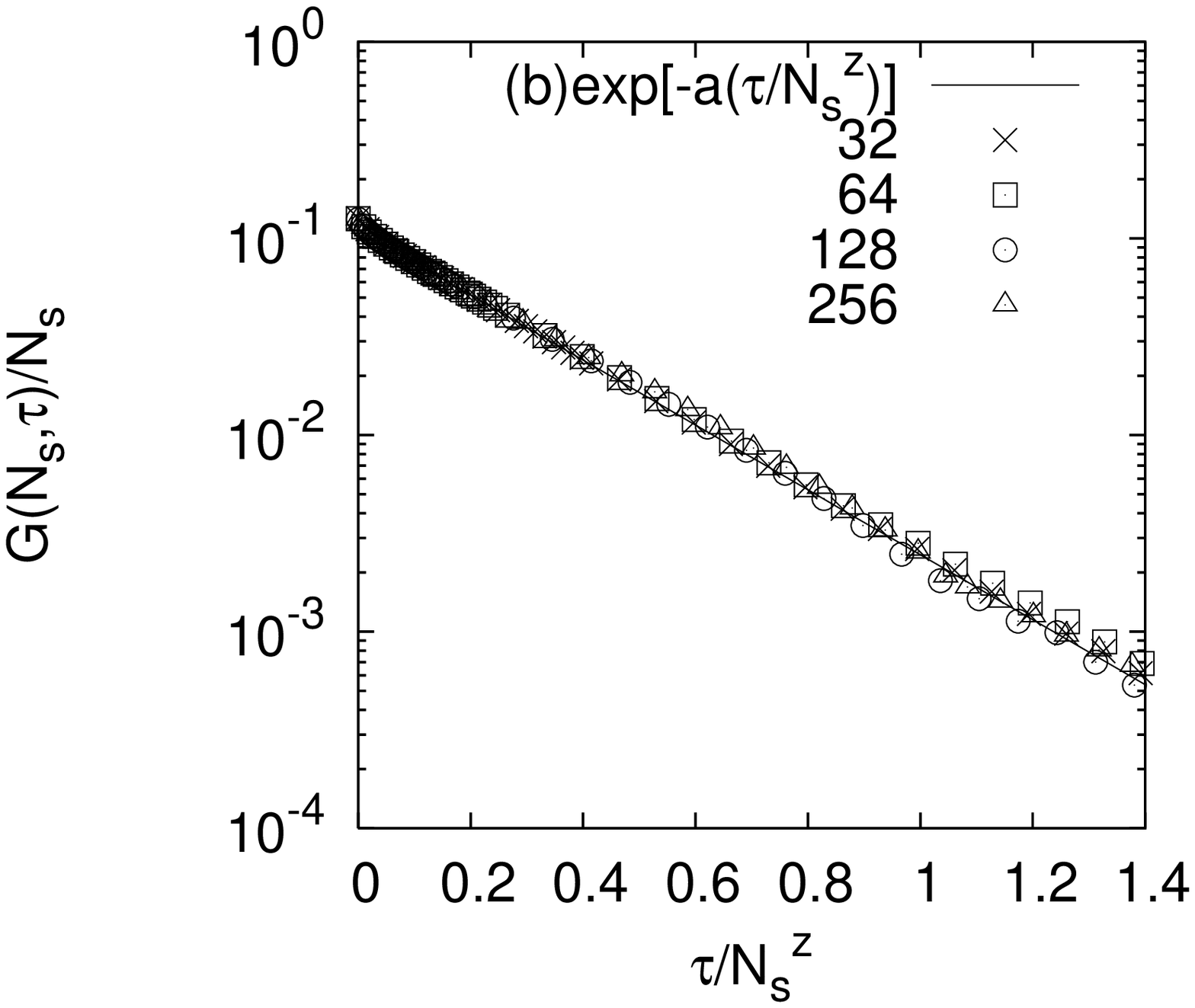}}
\caption
{The parked car correlator $G(N_s,\tau)$ in the MC phase,
for small $\tau$~(a), it decays linearly as function of 
$\tau^{1/z}/N_s$, and 
for large $\tau$~(b), it exponentially as function of $\tau/N_s^z$, 
where $z=3/2$. 
The data are obtained at $\rho_o=0.75$ and $\alpha=1$.}
\label{G_MC}
\end{figure}

This estimate fails to take into account finite-size effects.
Correlations are truncated near the two road edges
(all information is entering the garage).
The loss term is of the order
\begin{equation}
2 \int_0^{l} dx_0 ~(l-x_0) \tau^{2(\chi-1)/z} \sim l^2~ l^{2(\chi-1)}
\sim\tau^{2\chi/z}.
\end{equation}
This suggests that $G$ is of the form
\begin{equation}
G= N_s \left(  a -b \frac{\tau^{1/z}}{N_s}+\cdots \right),
\end{equation}
with constants $a$ and $b$, and
suggests that the scaling function ${\cal F}(\phi)$ is analytic 
at short times in the parameter $\phi=\tau^{1/z}/N_s$ 
instead of $\phi^{\prime}=\tau/N_s^z$.
Our numerical results shown in Fig.~\ref{G_MC}(a) 
are consistent with this.

In the opposite limit, $\phi,\phi^{\prime}\to \infty$, 
where time is large compared to the length of the road, 
all correlation cones are limited and equal to $l\simeq N_s$, 
and $G$ behaves as in Eq.~(\ref{g-fss-exp}), such that
\begin{equation}
G\sim N_s \times l \times g(0,\tau;\frac{1}{2}N_s,N_s)
\sim N_s ~e^{-a(\tau/N_s^{z})}.
\end{equation}
Our numerical results in Fig.~\ref{G_MC}(b)
are consistent with this as well.

\subsection{Inside the $C$ phase}

In the $C$ phase, the fluctuations scale with the same exponents as
in the MC phase, but the nonzero group velocity $v_g=1-2\rho_r$
changes completely the appearance of correlation functions, 
like $G(N_s,\tau)$. The density-density correlations spread just like 
in the MC phase, but only with respect to the moving frame of reference
and with $r$ replaced by $\tilde{r}=r+v_g\tau$.
The correlation function $g(r,t)$ scales as a power law 
$g\sim \tau^{2(\chi-1)/z}$ at $\tilde{r}=v_g\tau$ (i.e., $r=0$) 
but exponentially at nonzero $r$.
The fluctuation cones, $l \sim t^{1/z}$, are slanted and 
move with the flow.

\begin{figure}
\centerline
{\epsfxsize=6cm \epsfbox{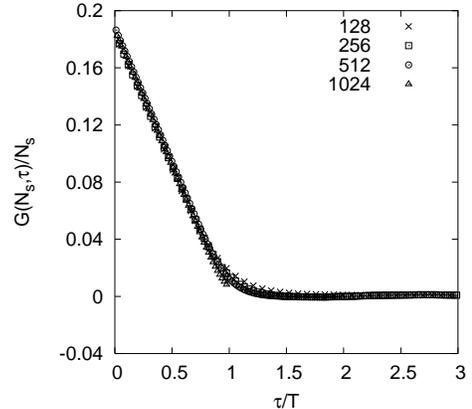}}
\caption
{The parked car correlator $G(N_s,\tau)$ in the $C$ phase;
it decays linearly as function of $\tau$ and becomes zero
after one time of flight $T=N_s/v_g(=\tau_{\rm flight})$.
Since it grows linearly with system size $N_s$, 
we plot $G/N_s$. The data are obtained at $\rho_o=0.75$ and
$\alpha=0.25$.}
\label{G_C}
\end{figure}

Figure~\ref{G_C} shows $G(N_s,\tau)$ inside the $C$ phase 
as a function of time $\tau$ for various road sizes $N_s$. 
It decays linearly until hitting zero at $\tau_{\rm flight}=N_s/v_g(\equiv T)$,
and then it remains at zero. All fluctuations move with the group velocity 
$v_g$ to the right and reach the garage at a constant rate. 
After one time of flight, $T$, 
all correlations with the initial configuration have disappeared.
The rounding in $G$ at $T$ is of order $\Delta t \sim T^{1/z}$,
and is due to the broadening of the remaining correlation cones
just before they are absorbed by the garage.

\subsection{Inside the $N$ phase}

\begin{figure}
\centerline
{\epsfxsize=6cm \epsfbox{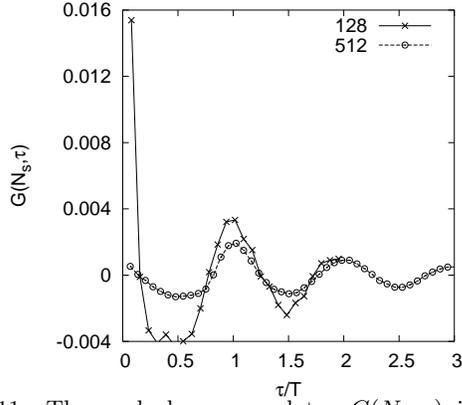}}
\caption
{The parked car correlator $G(N_s,\tau)$ in the $N$ phase;
the garage is no longer macroscopically occupied, 
such that $G$ does not scale with $N_s$ any more.
The correlations light up, like a lighthouse beam, 
at every time-of-flight interval $\tau_{\rm flight}=T$.
The data are obtained at $\rho_o=0.25$ and $\alpha=1$.}
\label{G_N}
\end{figure}

In the $N$ phase, the fluctuations in the total number of cars on the road,
$G(N_s,0)$, are not proportional to $N_s$ but are only of order one.
The garage is not macroscopically occupied any more and
acts very much like an ordinary road site.
Figure~\ref{G_N} shows the behavior of $G(N_s,\tau)$ inside the $N$ phase.
The total number of cars in the system is conserved, such that
$G$ reduces to
\begin{equation}
G(N_s,\tau)= \left<N_p(t_0+\tau) N_p(t_0)\right> -\overline N_p^2,
\end{equation}
and behaves similar to the autocorrelator $g(0,\tau)$.
The group velocity is nonzero, and therefore $G$ decays 
exponentially fast. However, because of PBC,
$G$ comes back to live, like a lighthouse light beam, 
after every time-of-flight interval $T(=N_s/v_g)$, 
with an amplitude of order $T^{2(\chi-1)/z}$
and with a temporal width of order $l\sim T^{1/z}$.

\subsection{At the $C$-$N$ transition}

\begin{figure}
\centerline
{\epsfxsize=6cm \epsfbox{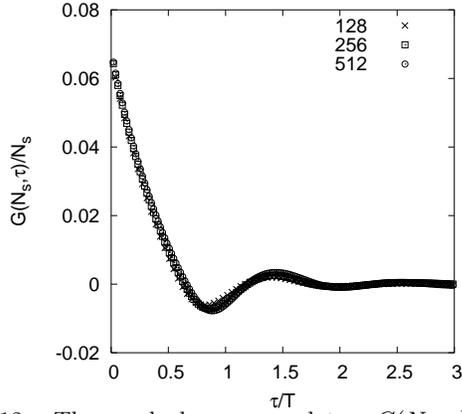}}
\caption
{The parked car correlator $G(N_s,\tau)$ at the $C$-$N$ transition;
it has a strong anticorrelation after one time-of-flight interval,
$T$, even though for small $\tau$ it decays linearly just like 
in the $C$ phase (and with almost the same group velocity). 
The data are obtained at $\rho_o=\alpha=0.25$.}
\label{G_CN}
\end{figure}

Figure~\ref{G_CN} shows how $G(N_s,\tau)$ at the $C$-$N$ transition 
decays in time for various system sizes. 
At small $\tau$, 
it decays linearly, rather like in the $C$ phase,
but then it seems to oscillate 
with a period determined by the time-of-flight time scale $T$;
$G(N_s,\tau)$ goes through zero at about $t\simeq \frac{1}{2} T$ 
and shows a strong anticorrelation at $t\simeq T$ 
(the maximum lies just before it).

\begin{figure}
\centerline
{\epsfxsize=6cm \epsfbox{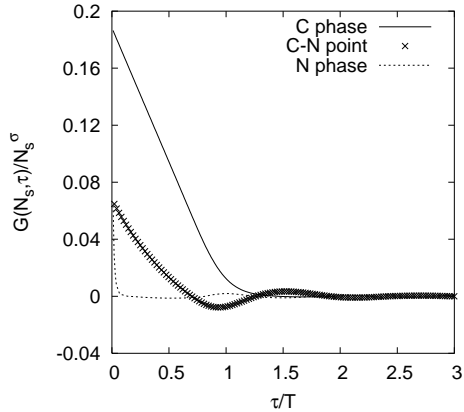}}
\caption
{Evolution of the parked car correlator $G(N_s,\tau)$ 
through the phase boundary at fixed system size $N_s=512$
along a line of constant group velocity: 
$v_g=1-2\rho_r$ with $\rho_r=0.25$;
$\sigma=1$ in the $C$ phase and at the $C$-$N$ transition,
while $\sigma=0$ in the $N$ phase.}
\label{plotG}
\end{figure}

Figure~\ref{plotG} illustrates how the transmission of information 
through the garage commences at the phase transition.
Suppose we approach the transition point from the $C$ phase 
following a line of constant $\alpha$. In the $C$ phase, 
these lines coincide with lines of constant group velocity. 
(In the $N$ phase, $v_g$ is constant along lines of constant $\rho_o$.)
So nothing changes on the road until we hit the transition point,
and $G$ decays linearly to zero and remains zero after one time-of-flight 
time scale. At the transition point, 
$G$ transforms abruptly into the oscillatory shape,
with an anticorrelation after one time of flight.
After that, it reduces inside the $N$ phase
to the lighthouse shape, in which $G(N_s,0)$ oscillates in phase
and does not scale with $N_s$ any more.

\begin{figure}
\centerline
{\epsfxsize=4.2cm \epsfbox{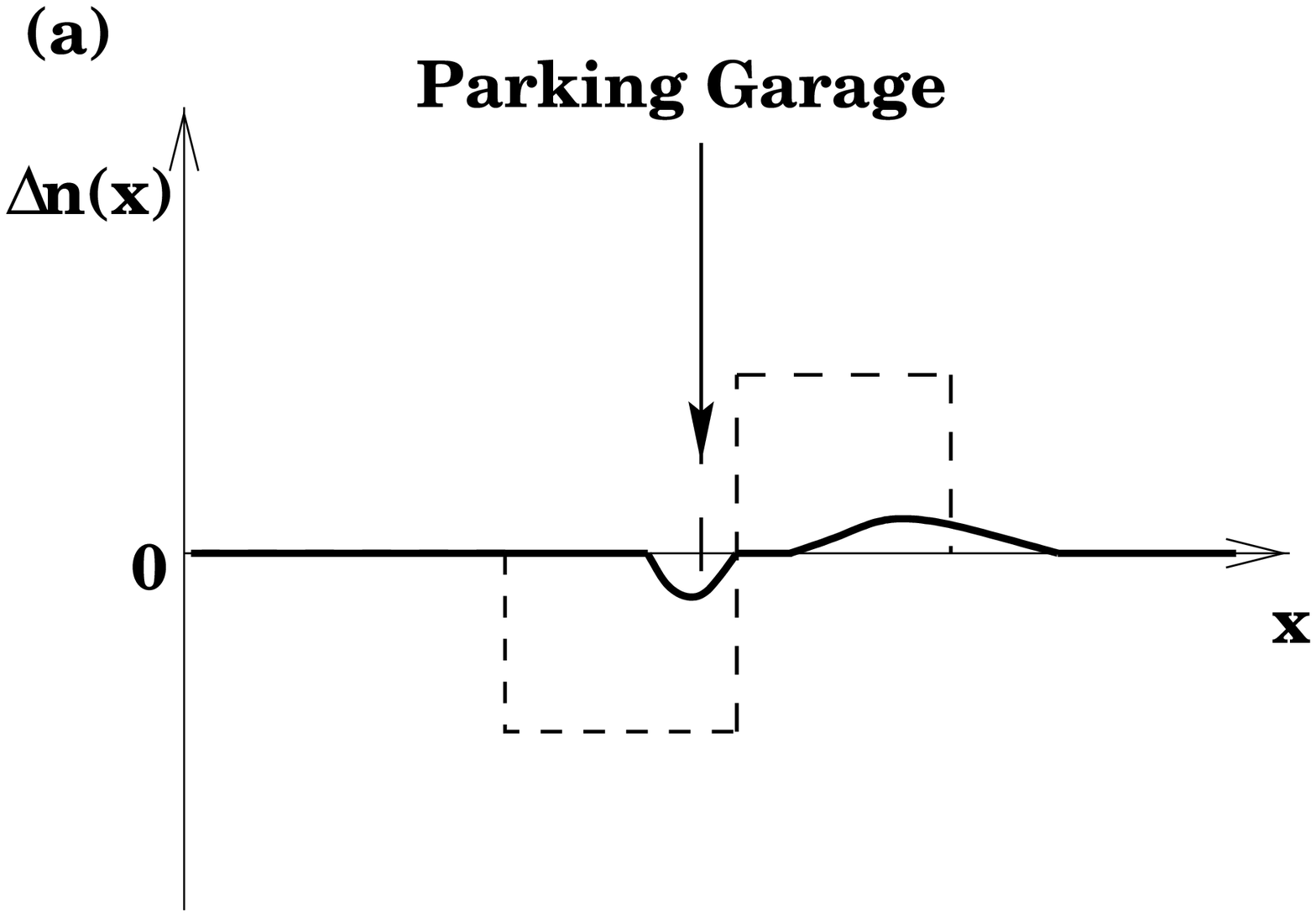}}
\centerline
{\epsfxsize=4.2cm \epsfbox{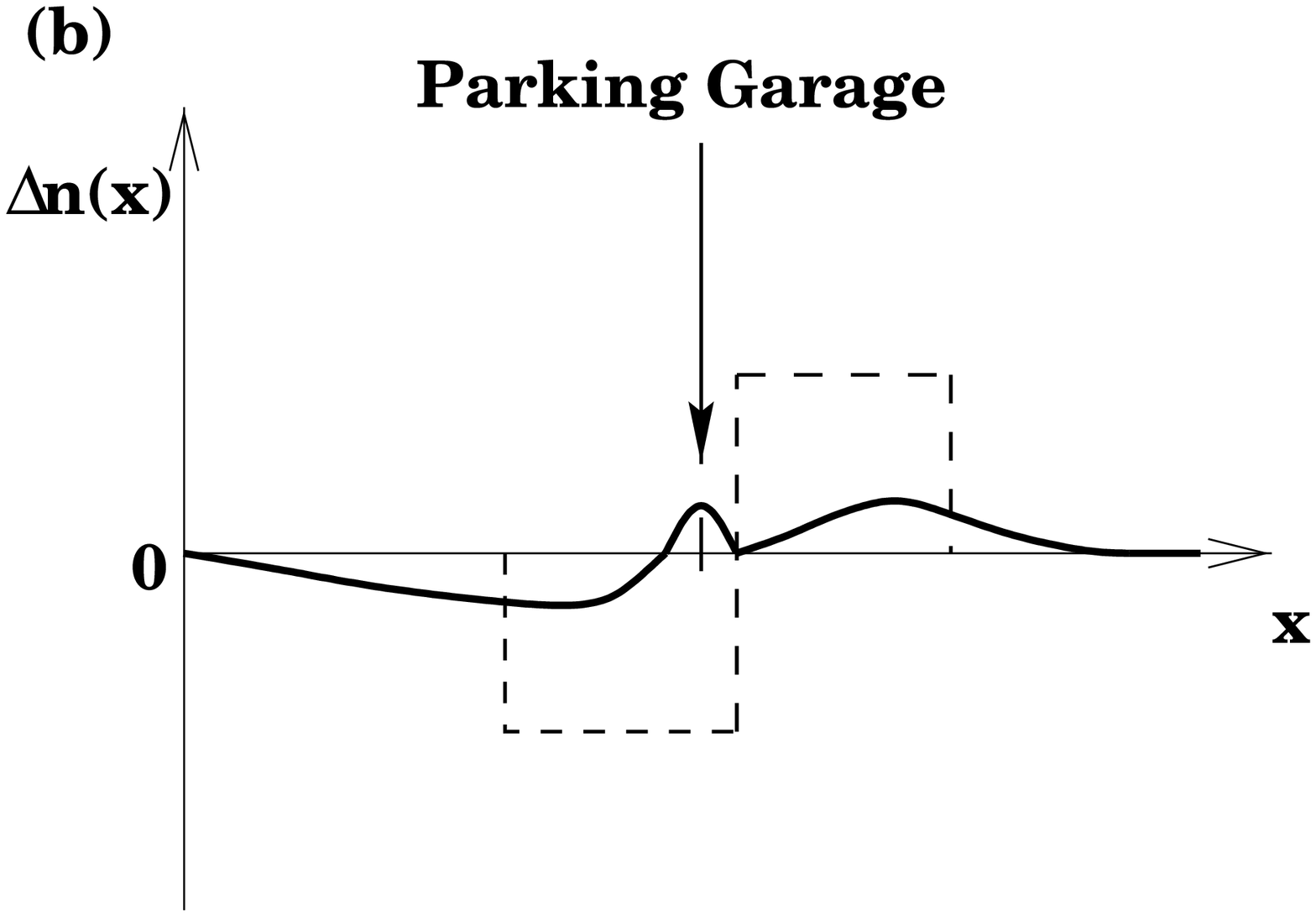}}
\centerline
{\epsfxsize=4.2cm \epsfbox{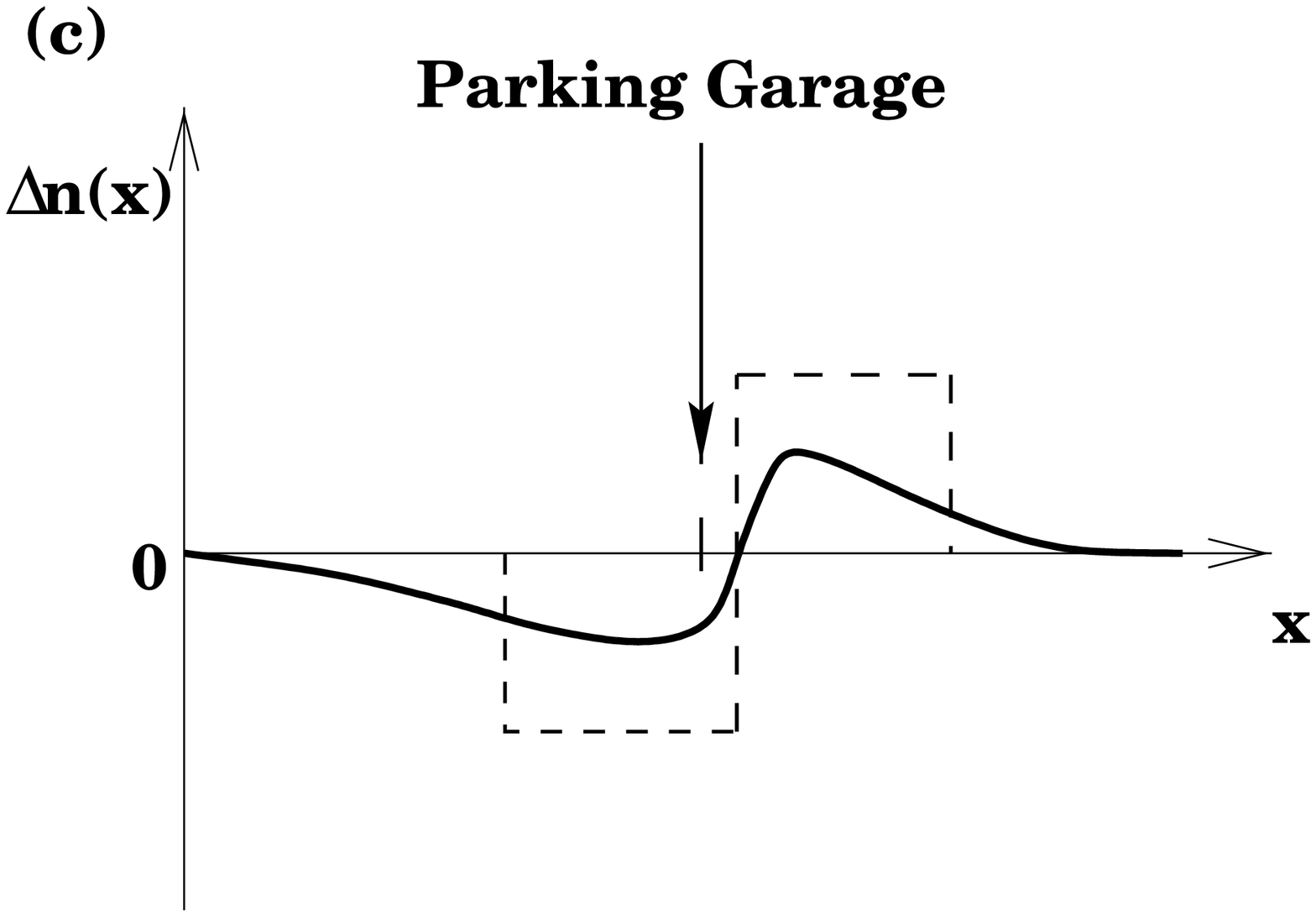}}
\caption
{Schematics of soliton-type local perturbations
diffusively spreading from the initial configurations (dashed lines) to
one time-of-flight interval later (solid lines);
(a)~in the $C$ phase,
(b)~at the $C$-$N$ transition, and
(c)~in the $N$ phase.}
\label{soliton}
\end{figure}

The anticorrelations at the $C$-$N$ transition point are intriguing
and need to be explained. Imagine a localized positive density fluctuation
at the beginning of the road at time $t_0$.
In the MC phase, it simply sits there 
while broadening as $l\sim\tau^{1/z}\sim\tau^{2/3}$
and weakening in amplitude as $\tau^{2(\chi-1)/z}\sim \tau^{-2/3}$.
In the $C$ phase, it broadens and weakens in the same manner, 
but travels like a solitary wave to the right with velocity 
$v_g=1-2\rho_r$ and drops out of the road after one time-of-flight
unit $T$. In the $N$ phase, it behaves very much the same, 
except that the positive density fluctuation creates a deficit 
inside the garage since the number of cars in the garage is finite, 
such that fewer cars can be put on the road
in the immediate wake of the positive fluctuation. Therefore, 
in the $N$ phase, every positive fluctuation carries a compensating 
(again localized) negative tail with it. Figure~\ref{soliton} illustrates 
the difference schematically, and our numerical simulations confirm 
this picture.

One could say that in the $N$ phase positive and negative local
excitations are bound in pairs, and that they unbind 
at the $N$-$C$ transition. On approach of the transition 
from the $N$ side, the width of the negative tail grows, 
but with conserved total area (equal to the area of the positive part 
of the excitation), because the average number of cars in the garage 
increases towards the transition (and diverges) and therefore
the reduced car output is being spread over more time.
At the transition point itself, the negative tail has vanished,
except for the finite-size scaling effect of order
$(N_s)^{-1/2}$.

Local excitations, thus, behave quite interestingly.
However, they do not explain the difference in the behavior 
of $G$ at the transition point and in the $C$ phase.
These solitary waves do not deplete the garage sufficiently 
to trigger its bottom because at the transition point 
the number of parked cars still diverges as $\sqrt{N_s}$.

Only nonlocal excitations, which encompass the entire system, 
are able to empty out the garage. Consider an excitation 
where the density of cars is globally and uniformly enhanced 
along the entire road, $\rho_i=\rho_r+\Delta$.
This requires that the number of cars to be taken out of the garage
should be proportional to the road length $N_s$. In the $C$ phase, 
those will not deplete the garage 
because the number of parked cars is also proportional to $N_s$.
Such a $\Delta$ regiment of extra cars marches with group velocity
$v_g$ to the right, reaches the end of the road, and thus returns to
the garage at a rate uniformly in time, row by row. 
Throughout this process, the garage is not aware of 
the existence of the regiment since we did not hit its bottom 
and because the information cones $l\sim\tau^{1/z}$ on the road 
do not broaden fast enough compared to the group velocity to maintain 
its memory at the garage exit. This implies 
that the exact stationary state rebuilds itself in the wake of the regiment,
and also that the enhanced road density decays linearly in time and 
vanishes completely after one time of flight $T$, just like our $G$ 
as a function of time $\tau$.

At the transition point, the garage only contains $N_p\sim \sqrt N_s$
cars, so that the same type of global uniform excitation can only 
have an amplitude of order $(N_s)^{-1/2}$, and always depletes the garage. 
This regiment of cars travels to the right with velocity $v_g$, just like 
in the $C$ phase, but in its wake the garage cannot rebuild 
the stationary state because it is empty. 
A depleted road density is established in the wake of 
the enhanced excitation, and thus anticorrelations build up,
and after one time of flight the average density of cars on the road is 
below normal (and a surplus of cars resides in the garage).
This explains qualitatively the oscillatory behavior of $G$ 
at the transition point and the anticorrelations at $T$.

\subsection{At the MC-$N$ transition}

\begin{figure}
\centerline
{\epsfxsize=4cm \epsfbox{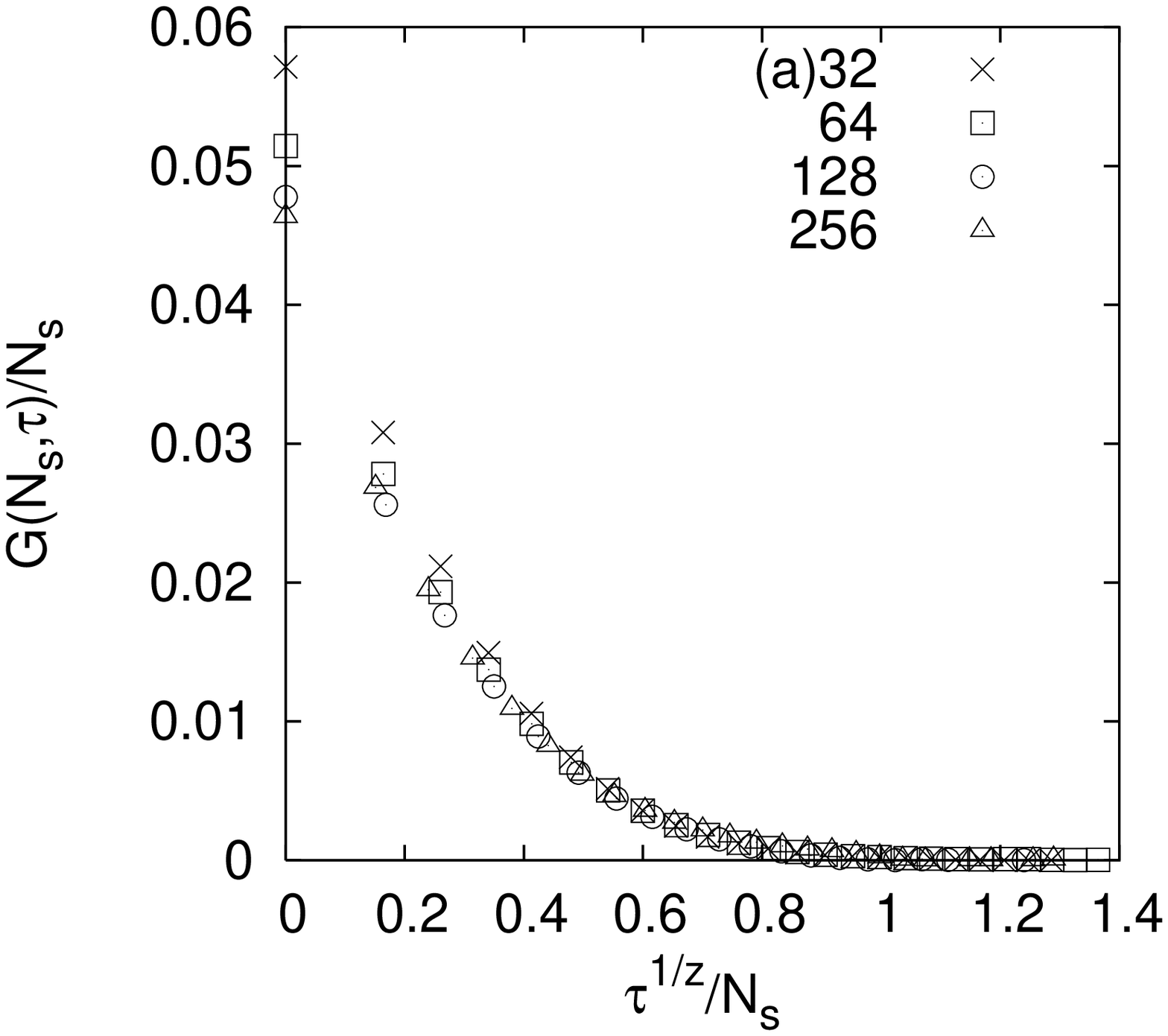} 
\epsfxsize=4.25cm \epsfbox{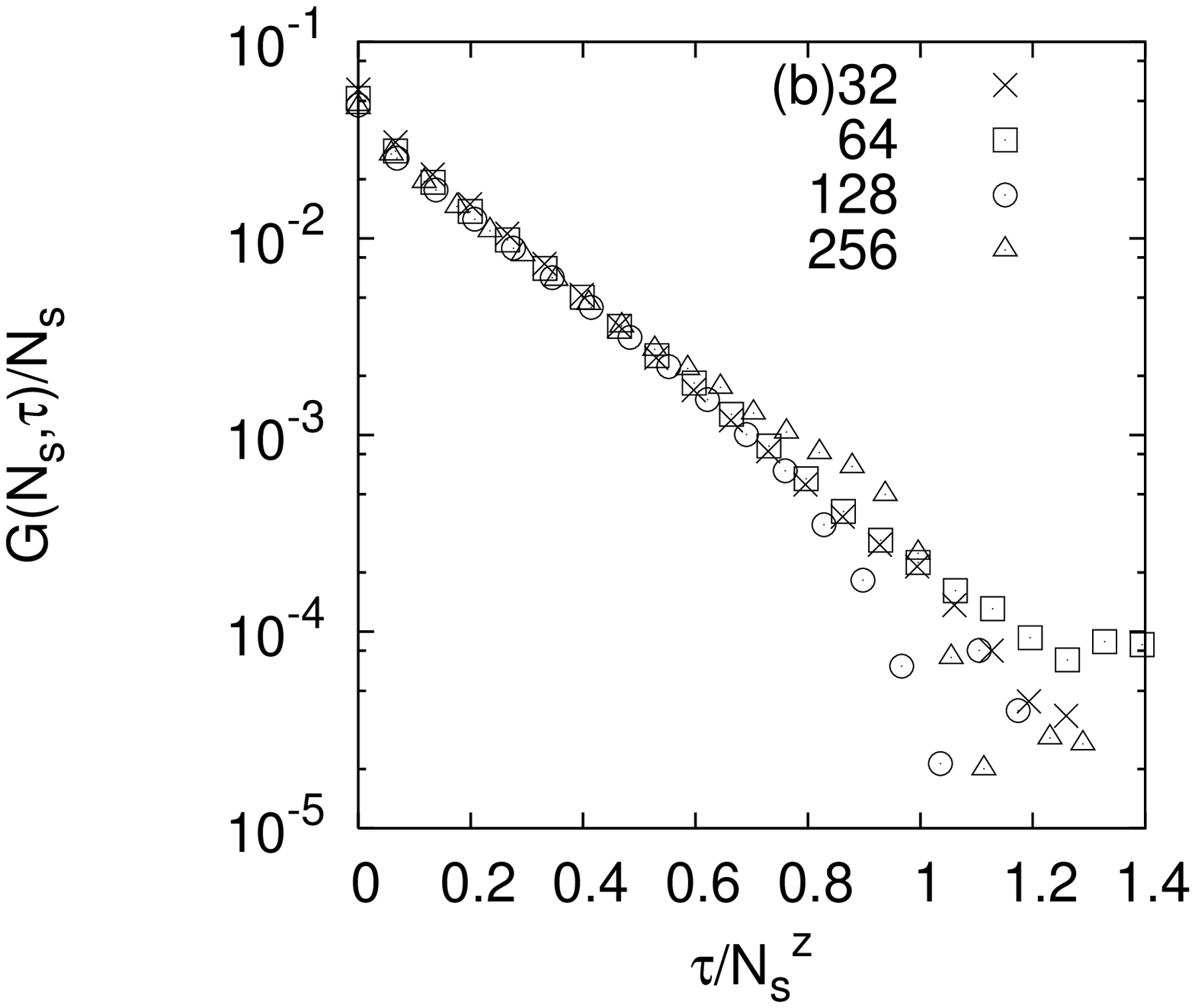}}
\caption
{The parked car correlator $G(N_s,\tau)$ at the MC-$N$ transition;
the shape is very similar to that inside the MC phase, see Fig.~\ref{G_MC}. 
The data are obtained at $\rho_o=1/2$ and $\alpha=1$.}
\label{G_NMC}
\end{figure}

The correlations at the MC-$N$ transition are 
less spectacular than at the $C$-$N$ transition.
The group velocity is zero inside the MC phase, and
still remains zero at the MC-$N$ transition.
Only inside the $N$ phase, does it start to shift continuously 
away from zero. Figure~\ref{G_NMC} shows how $G$ scales 
at the $MC-N$ transition. These numerical results are 
almost the same as those inside the MC phase
in Fig.~\ref{G_MC}.

Recall the argument about the behavior of $G(\tau, N_s)$ 
inside the MC phase (in section~{\bf IX~A}),
and imagine how this was modified at the MC-$N$ transition.
The factor $N_s$ in Eq.~(\ref{timezero}) represents the number of
cars on the road (the number of sites $x_0$ that are occupied). This
should be modified to $\rho_o N_s-a\sqrt N_s$, since the number of
parked cars scales as $\sqrt N_s$. The other terms, 
the spreading in time of the correlation cones
and the autocorrelations on the road, are likely unchanged.
Such differences are subtle and not surprisingly numerically invisible.

\section{Summary and Conclusion}
\label{summary}

In this paper, we presented the scaling properties of 
dynamic condensate phase transitions in terms of 
an 1D asymmetric exclusion process with a parking garage. 
There are two types of condensate
phases: the maximal current (MC) phase, where the road controls the
density of cars on the road, and the condensate ($C$) phase, 
where the garage (as a reservoir) controls the number of cars on the road.
The existence of a group velocity is
crucial for understanding the behaviors of correlations
and the density profiles in these two phases and at the phase transitions.

At both condensate transitions, the number of parked cars scales as 
$N_p\sim N_s^{y_p}$ with $y_p=\frac{1}{2}$,
while on approach of the transition,
the density of parked cars vanishes linearly with the control parameters
(the total density of cars in the system 
and the exit probability from the garage) 
$\rho_p\sim |\epsilon|^{\beta}$, with $\beta=1$.
Also, the transition points represent the onset of communication  
of information through the garage. This leads to interesting 
autocorrelations in the number of parked cars, 
particularly at the $C$-$N$ transition, 
due to the nonzero group velocity and associated time-of-flight effects.

Our parking garage model is a bare-bone version 
of dynamic Bose condensation and of queuing phenomena, like traffic jams.
The fundamental issue that needs further study is whether
the above scaling behavior, in particular the values of the critical
exponents, are universal or not.
How do more realistic interactions between the cars on the road
change this? Do the simplifications in the traffic jams, like
a stationary truck versus a moving one and ignoring the
spatial structure inside the queue of cars behind it by
collapsing them (piling them up) into a ``garage,"
affect the exponents?

There is some evidence suggesting 
that the exponents are indeed robust, e.g., 
in the Janowsky and Lebowitz~\cite{J-Lebowitz} model, 
the same simple KPZ-type values of the exponents appear 
in the fluctuations of the queue,  
although a detailed study of the queuing transition itself 
needs still to be performed. In addition, introducing short-range 
car-car interactions~\cite{Krug,interaction} do not seem to change 
the exponents either~\cite{ours}.

\section*{Acknowledgments}

We thank Joachim Krug for helpful discussions.
This research was supported by the National Science Foundation
under Grant No. DMR-9985806.

\end{multicols}
\end{document}